\documentclass[
  journal=largetwo,
  manuscript=research-article,
  year=2025,
  volume=00,
]{cup-journal}

\usepackage[nopatch]{microtype}
\usepackage{booktabs}
\usepackage{aas-macros}
\usepackage{epstopdf}
\usepackage{caption}
\usepackage{subcaption}
\usepackage{amsmath,amssymb}
\usepackage{gensymb}
\usepackage{longtable}
\usepackage{pdflscape} 
\usepackage{threeparttable}
\usepackage{multicol}
\usepackage{comment}
\usepackage{hyperref} 
\usepackage[range-phrase = \text{--}]{siunitx}
\usepackage{float}
\usepackage[export]{adjustbox}

\usepackage[australian]{babel}
\usepackage{csquotes}
\hypersetup{colorlinks,citecolor=blue,linkcolor=blue,urlcolor=blue}
\DeclareSIUnit{\yr}{yr} 
\DeclareSIUnit{\pc}{pc} 

\def\dmu{\ensuremath{\rm cm^{-3}\,pc}}

\def\rmu{\ensuremath{\rm rad\,m^{-2}}}
\def\TBhr{\ensuremath{\rm TB\,h^{-1}}}

\def\us{\ensuremath{\mu{\rm s}}}  

\def\deg{\ensuremath{{^{\circ}}}}

\def\sqdegphr{${\rm deg ^ {2} \, h^{-1}}$}

\def\zdm{\ensuremath{z_{\rm DM} }}
\def\zrm{\ensuremath{z_{\rm RM} }}

\newcommand{\vcsbeam}{{\sc vcsbeam}}
\newcommand{\presto}{{\sc presto}}
\newcommand{\dspsr}{{\sc dspsr}}
\newcommand{\psrchive}{{\sc psrchive}}
\newcommand{\birli}{{\sc birli}}
\newcommand{\hyperdrive}{{\sc mwa\_hyperdrive}}
\newcommand{\clfd}{{\sc clfd}}
\newcommand{\casa}{{\sc casa}}
\newcommand{\pint}{{\sc pint}}
\newcommand{\tempotwo}{{\sc tempo2}}
\newcommand{\nextflow}{{\sc nextflow}}
\newcommand{\psrpoppy}{{\sc psrpoppy}}
\newcommand{\spinifex}{{\sc spinifex}}

\newcommand{\paz}{{\tt paz}}
\newcommand{\pam}{{\tt pam}}
\newcommand{\pazi}{{\tt pazi}}
\newcommand{\psrzap}{{\tt psrzap}}
\newcommand{\pdmp}{{\tt pdmp}}
\newcommand{\rfifind}{{\tt rfifind}}

\newcommand{\psrfits}{{\sc psrfits}}
\newcommand{\vdif}{{VDIF}}

\newcommand{\snr}[0]{$\mathrm{S}/\mathrm{N}$}

\newcommand{\addref}[1]{\textcolor{red}{\textbf{(REF)}}}
\newcommand{\addurl}[1]{\textcolor{red}{\textbf{(URL)}}}

\defcitealias{pasaone}{Paper~I}
\defcitealias{pasatwo}{Paper~II}
\defcitealias{pasathree}{Paper~III}

\newcommand{\rev}[1]{#1}
\newcommand{\revtwo}[1]{#1}

\makeatletter
\DeclareCiteCommand{\citetalias}
  {\usebibmacro{prenote}}
  {\usebibmacro{citeindex}%
   \printtext[bibhyperref]{\@citealias{\thefield{entrykey}}}}
  {\multicitedelim}
  {\usebibmacro{postnote}}

\DeclareCiteCommand{\citepalias}[\mkbibparens]
  {\usebibmacro{prenote}}
  {\usebibmacro{citeindex}%
   \printtext[bibhyperref]{\@citealias{\thefield{entrykey}}}}
  {\multicitedelim}
  {\usebibmacro{postnote}}
\makeatother

\title[The SMART pulsar survey]{The Southern-sky MWA Rapid Two-metre (SMART) pulsar survey---IV. Survey update and an atlas of \rev{205} non-recycled southern pulsars}

\author{N.~D.~R.~Bhat}
\affiliation{International Centre for Radio Astronomy Research, Curtin University, Bentley, WA 6102, Australia}
\author{C.~P.~Lee}
\affiliation{International Centre for Radio Astronomy Research, Curtin University, Bentley, WA 6102, Australia}
\author{S.~J.~McSweeney}
\affiliation{International Centre for Radio Astronomy Research, Curtin University, Bentley, WA 6102, Australia}
\alsoaffiliation{SKA Observatory, SKA-Low Science Operations Centre, 26 Dick Perry Avenue, Kensington, WA 6151, Australia}
\author{B.~W.~Meyers}
\affiliation{Australian SKA Regional Centre (AusSRC), Curtin University, Kent Street, Bentley, WA 6102, Australia}
\alsoaffiliation{International Centre for Radio Astronomy Research, Curtin University, Bentley, WA 6102, Australia}
\author{C.~M.~Tan}
\affiliation{International Centre for Radio Astronomy Research, Curtin University, Bentley, WA 6102, Australia}
\author{M.~Xue}
\affiliation{National Astronomical Observatories, Chinese Academy of Sciences, Datun Road, Chaoyang District, Beijing 100101, China}
\author{Qiuyang~Fu}
\affiliation{National Astronomical Observatories, Chinese Academy of Sciences, Datun Road, Chaoyang District, Beijing 100101, China}
\author{N.~A.~Swainston}
\affiliation{International Centre for Radio Astronomy Research, Curtin University, Bentley, WA 6102, Australia}
\alsoaffiliation{CSIRO Space and Astronomy, PO Box 1130, Bentley, WA 6102, Australia}
\author{S.~M.~Ord}
\affiliation{CSIRO Astronomy and Space Science, PO Box 76, Epping, NSW 1710, Australia}
\author{G.~J.~Sleap}
\affiliation{International Centre for Radio Astronomy Research, Curtin University, Bentley, WA 6102, Australia}
\author{S.~E.~Tremblay}
\affiliation{National Radio Astronomy Observatory, 1011 Lopezville Road, Socorro, NM 87801, USA}
\author{W.~van~Straten}
\affiliation{ Manly Astrophysics, 15/41-42 East Esplanade, Manly, NSW 2095, Australia}
\author{A.~Williams}
\affiliation{International Centre for Radio Astronomy Research, Curtin University, Bentley, WA 6102, Australia}
\author{C.~Di~Pientrantonio}
\affiliation{Pawsey Supercomputing Research Centre, Kensington, WA 6151, Australia}
\author{C.~J.~Harris}
\affiliation{Pawsey Supercomputing Research Centre, Kensington, WA 6151, Australia}
\author{P.~J.~Elahi}
\affiliation{Pawsey Supercomputing Research Centre, Kensington, WA 6151, Australia}


\begin{document}

\begin{abstract}
The Southern-sky MWA Rapid Two-metre (SMART) survey, which capitalises on the MWA's large field of view and voltage recording capability, is an ambitious effort to conduct sensitive searches for pulsars and fast transients in the 140--170\,MHz band. The novelty of voltage recording, long dwell times (4800 s) and the high-time and -frequency resolutions (100\,$\mu$s/10-kHz) exchange a large survey speed ($\sim 450$\,\sqdegphr) for high computational cost.
The survey covers the entire sky south of $+30^{\circ}$ in declination through a series of dedicated observing campaigns, accumulating nearly four petabytes of data.
The large volumes of data and the processing challenges at low frequencies necessitate data processing to be approached in multiple phases, and the initial searches focused on a first-pass (shallow) survey of parts of the skies, as reported in earlier papers in this series. These data are also processed for re-detections of hundreds of known pulsars in the southern sky, many of which are also the first detections at frequencies below 400\,MHz. 
\rev{
This paper is motivated by the need to address the inherent difficulties (for the wider community) in handling large amounts of voltage data and software/processing challenges for routine pulsar detections, and also by the fast-evolving landscape of the SKA Observatory (SKAO). 
With the construction and commissioning ramping up towards the full-scale SKA-Low, a low-frequency catalogue of detectable pulsars in the southern sky will prove to be a valuable reference for the science verification exercise. 
}
A growing sample of low-frequency pulsar detections and measurements will also prove invaluable in a variety of science applications including population studies, survey simulations and emission beam models, refining interstellar medium models for electron densities and the spatial distribution of turbulence, and also for forecasting the detection prospects and survey yield from pulsar surveys planned with SKA-Low. We also present an electronic catalogue of various data products,  including pulse profiles, time series and multi-channel folded archives, along with the measurements of dispersion and rotation measures, and mean flux densities for the detected pulsars, and this will be periodically updated as more detections flow on from the ongoing data processing. 
\end{abstract}


\section{Introduction}
\label{sec:intro}

The Southern-sky MWA Rapid Two-metre (SMART) survey is an ambitious effort that leverages the Voltage Capture System (VCS) and large field-of-view (FoV) 
of the Murchison Widefield Array \citep[MWA;][]{tingay2013,wayth2018} to undertake an all-sky survey for pulsars and fast transients in the 140--170\,MHz band.
It is the first large-scale pulsar survey in the southern sky in the frequency band of the low-frequency component of the SKAO (SKA-Low; 50--350\,MHz). With its long dwell times (4800\,s) and high-time and -frequency resolutions (100\,$\mu$s/10 kHz), the SMART survey will be much more sensitive than previous-generation southern-sky pulsar surveys \citep[e.g.,][]{70cm,manchester2001,keith2010}. The large data rates (0.5--1\,TB per min, or 42--84 TB per observation) and associated computational challenges necessitate data processing to be approached in multiple phases, starting with a first pass (i.e. shallow survey) as we reported in earlier papers \citep{pasaone}. 

\rev{
The survey data collection began in 2018, and through a series of dedicated campaigns 70\% of the sky was covered by mid-2021. 
There have been encouraging results 
from a partially-completed first-pass processing, including several new pulsar discoveries after a small fraction ($\sim$5\%) of the survey data analysed, and re-detections of 120 known pulsars from initial data quality checks and analysis \citep{pasatwo}.
}
Among the new discoveries are a variable sub-pulse drifter \citep{mcsweeney2022}, a pulsar that shows quasi-periodic nulling \citep{psrfive}, and low-luminosity steep-spectrum objects \citep{psrone}. SMART data have also been used to confirm pulsar candidates identified in image-based transient searches \citep{2025ApJ...981..143M}. With completion of the last SMART observation (November 2023), approximately 4\, PB of data have been accrued, thus marking an important milestone, and making it the largest all-sky survey in the southern sky, and the second largest (in data volume) after the LOFAR Tied-Array All-Sky (LOTAAS) survey \citep{sanidas2019}.

While the adoption of VCS  \citep{tremblay2015,mwaxref} makes longer-term data archiving tractable, the inherent complexities of calibrating and beamforming such voluminous data makes data processing rather tedious and computationally intensive, especially for routine pulsar detections and analysis. 
Indeed, voltage data offers enormous flexibilities through a multitude of reprocessing avenues, as newer and better algorithms and faster processing pipelines become available in the future. 
With an intent to better serve the wider pulsar community, and making a range of useful data products publicly accessible, we have therefore embarked on a systematic processing of SMART data sets to make a low-frequency southern-sky catalogue of pulsar detections. 
\rev{
With the SKAO construction ramping up and a rapid progression expected in the coming years\footnote{See \href{https://www.skao.int/en/science-users/timeline-science}{SKAO Timeline to Science} webpage.}, it is therefore timely and important to develop such a catalogue as a living resource.}
The MWA is also the Australian Precursor for SKA-Low, and AA1 (Array Assembly 1, with 18 stations)\footnote{SKA1 Design Baseline Description, \href{https://zenodo.org/records/16895574}{SKA-TEL-SKO-0001075}.} is expected to reach (or even surpass) the sensitivity of Phase III MWA. A comprehensive catalogue of southern pulsars detectable in the SKA-Low frequency band can thus be a valuable resource in the validation and science commissioning and verification of these  stations. 

As the MWA upgrade transitioned to Phase III, early prototypes were developed and commissioned for SKA-Low, providing opportunities for useful demonstrations. Over the past few years, we have taken some of the first steps in this direction, by developing pulsar capabilities for the early prototype stations for SKA-Low \citep[i.e., EDA2 and AAVS2;][]{edaref,aavsref}, and through their commissioning and successful early science demonstrations. An initial census of pulsars and spectral analysis was reported in \citet{lee2022}, and more recently, the commissioning of the upgraded system (with a recording bandwidth of 25\,MHz) enabled the first polarimetric validation as well as science application, specifically probing the magneto-ionic structure of the interstellar medium (ISM) towards the Vela pulsar \citep{lee2024}.


\revtwo{
The SMART survey is also  motivated by several generations of past surveys; e.g., Parkes multibeam surveys \citep{70cm,manchester2001,keith2010}. 
These predecessor surveys have proven very useful, both to guide and prepare the next-generation surveys, and as a reference to cross-check pulsar discoveries and detections. 
Re-detecting a new pulsar candidate in a past survey often yields an initial estimate of the period derivative and hence the surface magnetic field strength and characteristic age. With SKA-Low anticipated to take on a bulk of the survey load \citep[e.g.,][]{keane2015,keane2025}, it is imperative to have a comprehensive southern pulsar survey at low frequencies; the SMART survey will fill this gap. 
This motivated securing resources for longer term data archiving in addition to the MWA Data Archive, through Australia Research Data Commons (ARDC)\footnote{\href{https://researchdata.edu.au/mwa-smart-survey-2020-2021/2304894}{MWA SMART Survey Raw Data 2020-2021} (doi:10.25917/HNZA-1T90).}, under data sets of national significance.
}


Aside from these strategic perspectives, a low-frequency pulsar catalogue is scientifically motivated on several counts.
While  large surveys provide subsets of such pulsar samples, the advent of large-FoV instruments, such as Low Frequency Array (LOFAR), Canadian Hydrogen Intensity Mapping Experiment (CHIME) and the MWA, also provide the ability to sample large patches of the sky instantaneously, and thus efficiently monitor or detect multiple pulsars in a single observation. 
\revtwo{
Large samples of pulsars and their measurements are very useful in understanding emission properties (e.g., beam shapes and luminosity distributions), especially when low-frequency data are combined with the published measurements at high frequencies (e.g., at 1-2 GHz) using MeerKAT \citep[e.g.,][]{Keith2024}.}
Measurements at low frequencies are also useful for dispersion and scattering/scintillation studies, especially for pulsars at low to moderate DMs, whereas it may be hard to discern such effects at higher frequencies ($\gtrsim$1\,GHz). 
These are important inputs for developing improved Galactic electron density models \citep{ne2001,ymw16,occ20,ne2025} and mapping out the spatial distribution of the turbulence strength.
\revtwo{An improved sample of low-frequency pulsar detections are also important for accurately modelling the detectable population  at low frequencies \citep{keane2015,xue2017,pasaone}, and thus inform future large surveys planned with SKA-Low.}

\begin{figure*}[t]
\begin{center}
\includegraphics[width=0.63\linewidth]{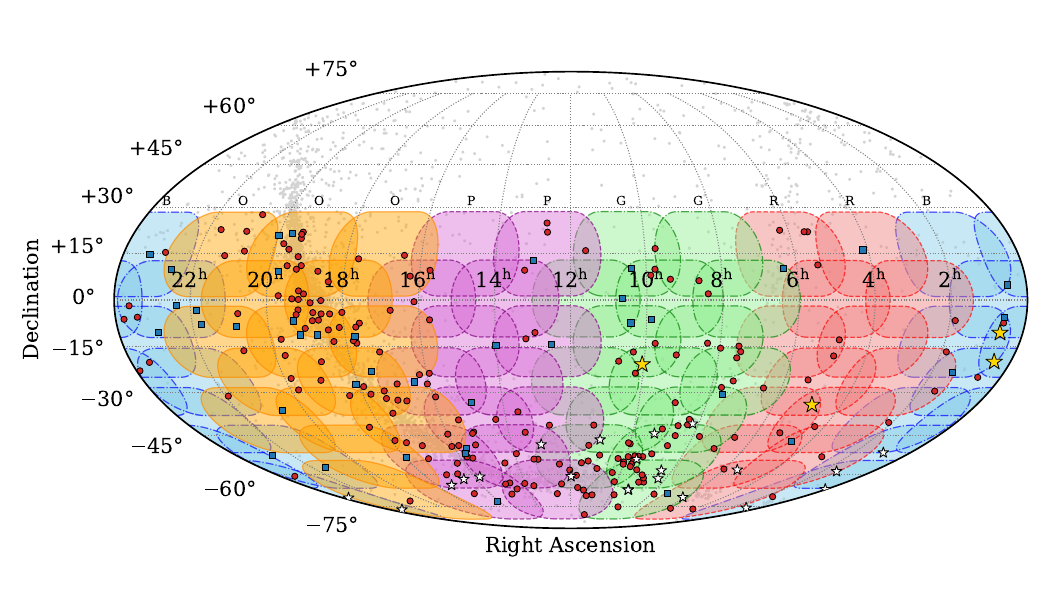}
\hfill
\includegraphics[width=0.36\linewidth]{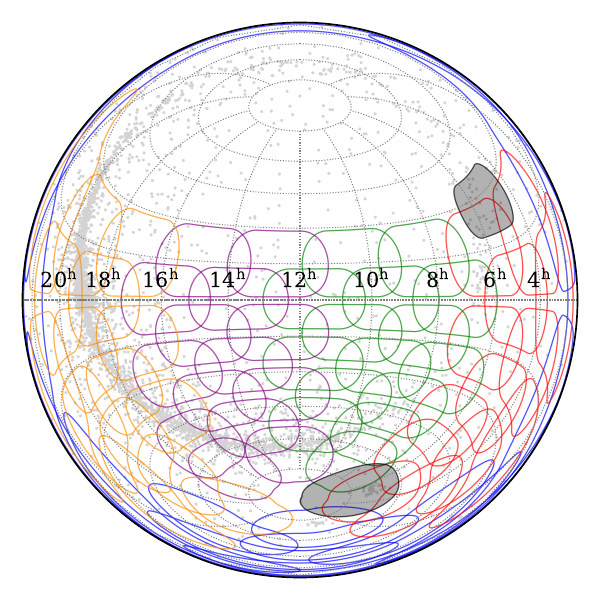}
\caption{
\rev{
Sky plots in Mollweide projection (left) and Lambert azimuthal equal-area projection (right) summarising the observing strategy adopted for the SMART pulsar survey and the progress made to date with data processing.}
\rev{
The full visible sky (i.e., declination $< +30\deg$) is covered in 71 pointings that overlap $10\deg$ in right ascension and $15\deg$ in declination.
The coloured contours represent the half-power points of the main lobe of the primary beam (at 154\,MHz), normalised by the maximum power in each beam.}
The blue, red, and green pointing sets, as well as 10 of the 14 purple ones, were observed through four dedicated observing campaigns undertaken in the 2018B, 2019B, 2020A, 2021A semesters.
The remaining four purple pointings and the orange pointing set were observed in the 2023A semester.
The last pointing, shown in black in the right plot, covers the South Celestial Pole (RA: 05h30m, declination $-90\deg$) and was observed in the 2023B semester.
\rev{
The top row of pointings are labelled by their colour to indicate the pointing set that they belong to.
The red filled circles are the known non-recycled pulsar detections (205 so far), and the blue squares are the known millisecond pulsar detections \citepalias[40 so far, as reported by][]{pasathree}.
The yellow stars show the pulsar discoveries from the first-pass (pilot) search processing of parts of the sky (see \citetalias{pasaone} and \citetalias{pasatwo} for further details), and the smaller white stars show the unpublished discoveries from the ongoing (second-pass) search processing.
The grey dots show all known pulsars from the ATNF pulsar catalogue \citep[v2.7.0;][]{manchester2005}.}
}
\label{fig:smartprogress}
\end{center}
\end{figure*}

\revtwo{In this sequel to the SMART survey papers \citep[][referred to hereafter as Papers I, II and III]{pasaone,pasatwo,pasathree}, we present an updated status of survey data collection and a more complete catalogue of pulsar detections thus far, along with useful data products including the pulse profiles, time series and measurements. 
A census of millisecond pulsars (MSPs) was presented in \citetalias{pasathree}. 
In Section~\ref{sec:datacollection}, we present a survey update, and in Section~\ref{sec:dataprocessing} we describe various steps involved in processing data from the upgraded VCS system (known as MWAX VCS), including calibration and beamforming.}
The pulsar data analysis, including mitigation of radio-frequency interference (RFI) and the measurements of pulsar and ISM properties, is described in Section~\ref{sec:dataanalysis}.
The catalogue and enhanced data products are summarised in Section~\ref{sec:catalogue}, some discussion and future work in Section~\ref{sec:discussion}, and a summary in Section~\ref{sec:summary}.


\begin{figure*}[t]
\begin{center}
\includegraphics[width=0.99\linewidth]{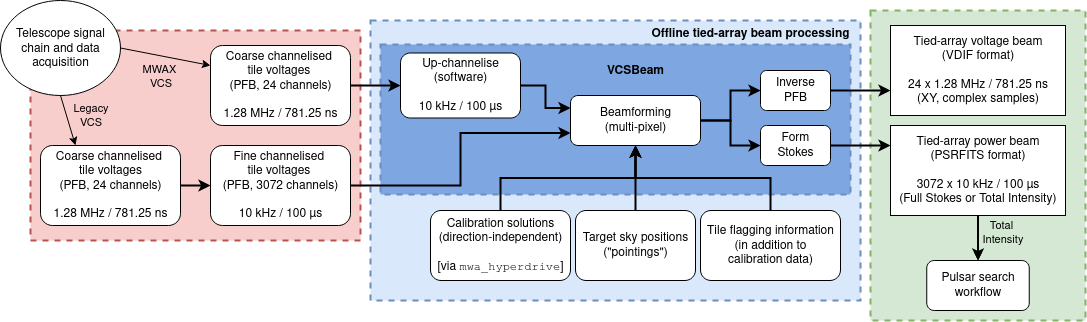}
\caption{
The signal path and processing workflow that makes the subsystem for the high-time resolution science observations with the MWA. With the legacy system (retired in August 2021), voltage data are recorded after two stages of polyphase filterbank (PFB) channelisation, resulting in 3072 $\times$ 10-kHz voltage time series at 100-$\mu$s resolutions, which can be then beamformed on targets of interest after the calibration procedures. The beamformed data are written out as detected full Stokes, or summed polarisation (for the SMART survey project) at the native 100-$\mu$s/10-kHz resolutions, or can be reprocessed using an inverse synthesis filter to produce a higher resolution ($\approx$0.78\,$\mu$s) time series (e.g., for MSP observations). With the new system (MWAX VCS; starting October 2021), voltage data are recorded directly after the coarse channelisation (24 $\times$ 1.28-MHz wide channels), which can be then processed (offline) in different ways. In the `SMART' mode, data in an identical format to the legacy system are generated after applying a fine PFB (in software), to result in 100-$\mu$s/10-kHz resolutions. For other science applications that require higher time resolution (e.g., MSPs) or visibility data (e.g., FRBs) the same data can be processed using suitable alternatives to the \vcsbeam\ suite of software tools. See text for further details. 
}
\label{fig:mwaxvcsbeam}
\vspace{-0.5cm}
\end{center}
\end{figure*}

\section{SMART Survey: Data Collection}\label{sec:datacollection}
\begin{table*}[t]
    \begin{threeparttable}
        \caption{Summary of the SMART survey observing campaigns.}
        \label{tab:smartobs}
        \begin{tabular}{lccccccc}
            \toprule
              \multicolumn{1}{l}{Observing } &
              \multicolumn{1}{c}{Dates of observation} &
              \multicolumn{1}{c}{No. of VCS} &
              \multicolumn{1}{c}{Pointing} &
              \multicolumn{1}{c}{RA range} &
              \multicolumn{1}{c}{Data collected} &
              \multicolumn{1}{c}{Re-detected}  &
              \multicolumn{1}{c}{New} 
            \\
              \multicolumn{1}{l}{semester} &
              \multicolumn{1}{c}{ } &
              \multicolumn{1}{c}{pointings} &
              \multicolumn{1}{c}{labels$^{\dagger\dagger} $} &
              \multicolumn{1}{c}{(hr)} &
              \multicolumn{1}{c}{(TB)} & 
              \multicolumn{1}{c}{pulsars$^{\dagger} $}  &
              \multicolumn{1}{c}{pulsars$^{\ddagger} $}  
            \\
            \midrule
            2018B & September -- December 2018 & 13 & B01--B13 & 22--3 & 546 & 24 (0) & 9 \\
            2019B & September -- December 2019 & 13 & R01--R13 & 2--7 & 546 & 29 (1) & 2 \\ 
            2020A & January -- March 2020 & 15 &  G01--G15 & 6--12 & 630 & 62 (6) & 10 \\
            2020B & March -- May 2021 & 10 & P01--P10 & 12--16 & 420 & 34 (0) & 1 \\
            2023A & April -- August 2023 & 4 & P11--P14 & 12--16 & 256 & 25 (0) & 3 \\
            2023A & April -- August 2023 & 15 & O01--O15 & 16--21 & 1280 &  80 (2) & \\
            2023B & November 2023 & 1 & SCP &  5--6 & 85 & 1 (1) & \\
            \bottomrule
        \end{tabular}
        \begin{tablenotes}
            \item[$\dagger$] The number in parentheses are common re-detections with the previous semesters.
            \item[$\ddagger$] New pulsars discovered thus far in the ongoing processing; unpublished
             ones will be reported in a future paper in the SMART series.
             \item[$\dagger\dagger$] See Table~\ref{tab:obsids} and Figure~\ref{fig:smartprogress} for details; the label ID corresponds to Blue, Red, Green, Purple and Orange pointings as shown in Figure~\ref{fig:smartprogress}, and SCP denotes the pointing towards the South Celestial Pole.
        \end{tablenotes}
    \end{threeparttable}
\end{table*}

As described in earlier papers, through a series of four dedicated campaigns from 2018 to 2021, we advanced the SMART survey data collection to 70\%, covering the sky within right ascension (RA) $<$16\,h and $>$21\,h and declination $\delta < +30^{\deg}$. The fourth campaign (2021A) was only partially complete before the upgrade activities commenced in earnest to build a new correlator for the MWA and implementing changes to the signal path, as part of the planned transition of the array to Phase III. The commissioning of the new MWAX correlator and VCS \citep{mwaxref}, and associated changes to the signal path and data format, necessitated a significant revamp in the sub-system designs that constitute the high time resolution (software) backend for processing VCS recorded data, as summarised in Section~\ref{sec:vcsbeam}.

For the legacy system, which comprised 128 tiles and performed fine channelisation (128 $\times$ 10 kHz) before recording (as 4+4 bit complex voltage samples), the associated data rate was 28\,\TBhr{}.
With the higher bit depth of the new system (from a 4-bit to 8-bit sampling), the data rate is twice as large. Furthermore, recording is done following the coarse (1.28-MHz wide) channelisation, as opposed to the post-fine channelisation stage for the legacy system. This also meant adapting to new data format, and the integration of the {\tt mwalib} package to flexibly handle data from both the legacy VCS system (until May 2021) and the new system MWAX VCS (August 2022 onward). Additionally, the installation of two new receivers as part of the ongoing Phase III integration resulted in recording additional 16 tiles (i.e. recording from all 144 tiles by default, in place of previously 128), and consequently, an increased data rate of 64\,\TBhr{}, which is a significant step up from the earlier 28\,\TBhr{}. The longer baselines of additional tiles, in principle, may offer the benefits of improved localisation for new pulsar discoveries, and therefore the final campaign employed recording from all 144 tiles. A summary of these SMART campaigns is given in Table~\ref{tab:smartobs}, along with details like the amount of data collected and the number of pulsars detected thus far. 

The final observing campaign was carried out in the 2023A observing semester, which covered the incomplete pointings from the 2021A (fourth) campaign (i.e. purple pointings) as well as the full sky in the $16 < \mathrm{RA} < 21$\,h and $\delta < +30^{\deg}$ (i.e. orange pointings) through 15 + 4 = 19 VCS pointings, the details of which are summarised in Table~\ref{tab:smartobs}. Additionally, a dedicated SMART observation was made towards the South Celesial Pole (SCP), as otherwise  
this part of the sky would end up with only partial low-sensitivity coverage. As can be seen from the survey sky map in Figure~\ref{fig:smartprogress}, there is $\sim$20\% overlap in the sky coverage with earlier campaigns (blue and purple pointings), and that is to do with the nature of our survey design, where every single patch of the sky is covered in at least two pointings at a gain above the half power (primary) beam level. The total data collected from the final campaign is 1621 TB from a total of 19 pointings, thereby bringing the net data collection from SMART to $\sim$3.9\,PB (from a total of 71 pointings). Further details of the SMART pointings including the coordinates of the pointing centres and observation IDs (Obs IDs) for VCS and calibrator observations are listed in Table~\ref{tab:obsids}. These data sets effectively represent a digital version of the entire sky south of $+30^{\deg}$, and serves as a unique permanent record of the electromagnetic state of the sky. 
\revtwo{
The native resolutions (for the legacy VCS data) are 100\,$\mu$s, 10\,kHz in time and frequency, respectively; however, these can be reprocessed for higher time resolution (down to $\sim1$\,$\mu$s) using a synthesis filter \citep{mcSweeney2020}. As such the SMART datasets hold an enormous legacy value, and recording as voltages (instead of beamformed data) makes their long-term archiving tractable.}\footnote{With a data rate of 138 GB per beam (for 4800 s), beamforming will result in a significant data explosion; e.g., 670 TB from a single observation when tessellated into 5000 beams across the $\sim$610\,${\rm deg^2}$ FoV.}
%


The successful completion of data collection marks a significant milestone for the SMART survey project. 
\revtwo 
{As described in earlier papers,  we first performed a \emph{shallow} survey of a small fraction ($\sim$5\%) of the sky, reaching approximately one-third of the full-search sensitivity}. This effectively served as a proof-of-concept. 
We emphasize that due to the drift-scan nature of SMART observations, the \emph{effective on-source time} will vary depending on the pointing and the offset relative to the beam centre, as discussed in \citetalias{pasaone} (cf. Section~2.5, Figures 3 and 4).
\revtwo{
The ongoing deep-pass processing involves processing the full 4800-s observations; this commenced in September 2024, resulting in the discovery of 23 new pulsars after processing $\sim$10\% of data.}
Follow-ups of these are in progress and will be reported in a future paper, where we will also describe the extended search pipeline and processing strategies.


\section{Processing MWAX VCS Data}\label{sec:dataprocessing}
The SMART project also strongly motivated the development and commissioning of the software subsystem \vcsbeam{} that was pivotal for survey completion. \vcsbeam{} was developed as a flexible software that can process data from both legacy and the new (MWAX) VCS systems. Furthermore, developments on the calibration side have enabled us to undertake a systematic processing of all calibration observations in a uniform manner.
As these new software and processing steps will become the standard for the routine processing of VCS data from here on, we will summarise the related procedures here, as a useful reference. 
\rev{
This involves obtaining calibration solutions using dedicated observations of calibrators (recorded in the visibility mode), and generating data in identical format to the legacy VCS system (i.e. 100-$\mu$s/10-kHz resolutions). 
An overview of the related signal path and processing workflow is presented in Figure 2.
}

\subsection{Phase and amplitude calibration}\label{sec:calsols}
For each of the SMART observations, there are a handful of standard calibration observations taken before and after the nominal voltage (VCS) recording.
These are required in order to derive the relative complex amplitudes and phases of each MWA tile (as a function of frequency channel), which are a critical part of the beamforming process. 
This collection of tile-based amplitudes and phases across the observing bandwidth are the `calibration solution.'
We transfer these calibration solutions from the standard calibrator field to the SMART observation field -- the MWA is generally phase-stable on reasonably long time scales, and our experience has been that the use of a single calibration solution works well over $\sim$4800-s observing durations of SMART pointings. 

To extract the calibration solution, we used the purpose-built \birli{}\footnote{\url{https://github.com/MWATelescope/Birli}} and \hyperdrive{}\footnote{\url{https://github.com/MWATelescope/mwa_hyperdrive}} packages on the dedicated calibration observation. 
First, \birli{} is used to flag known misbehaving tiles (based on metadata collected by the observatory during observing sessions), apply passband, digital gain and geometric corrections, perform basic radio frequency interference (RFI) mitigation, and average the visibilities to 2-s/40-kHz resolution. 
\rev{
These pre-processed visibilities are then passed to \hyperdrive{}, along with nominal sky and primary beam models, which attempts to solve for the amplitude and phase of each tile at each frequency channel and incorporates ionospheric refractive source shifting effects \citep{hyperdrive}, i.e., bulk source position offset correction across the field-of-view. 
}
This amounts to a direction-independent calibration procedure that is equivalent to an iterative application of the classic \textit{antsol} approach \citep[e.g.,][]{antsol,rts}. 
The calibration solutions obtained in this manner are collated at a central repository and can be used to expedite future re-processing. 
In general, there are two or more dedicated calibrator observations for each SMART observation, and the one that maximises the detection sensitivity of known bright pulsars in the field is adopted as the preferred calibration solution for a given observation.   

\subsection{Tied-array beamforming with VCSBeam}\label{sec:vcsbeam}
\rev{
The algorithm used for the beamforming operation 
in \vcsbeam{}\footnote{\url{https://github.com/CIRA-Pulsars-and-Transients-Group/vcsbeam}}
is identical to that described by \citet{ord2019}, with the addition of the multi-pixel functionality \citep{swainston2022} and the ability to invert the polyphase filterbank (PFB) to produce coarse-channelised VDIF (VLBI data interchange format) data \citep{mcSweeney2020}. }
To summarise, the basic beamforming operation involves:
\begin{enumerate}
    \item Applying a predetermined calibration solution to  complex-valued voltages, to account for differences between the responses and signal paths of each MWA tile and polarisation (Section~\ref{sec:calsols});
    \item Adding a term to the phases of the voltages to account for the geometric delay of a signal coming from the target look-direction;
    \item Summing the voltages from each tile/polarisation together; and
    \item (Optionally) converting the dual polarisation beamformed voltages to Stokes parameters.
\end{enumerate}

The advent of the MWAX backend removed the fine-channelisation (i.e. the second PFB stage) step that was previously performed by dedicated field programmable gate array (FPGA)  boards on site before the signal was passed to either the correlator or the VCS backend \citep[cf.][]{tremblay2015}. Consequently, MWAX voltage data are written to disk as coarse-channelised, Nyquist-sampled voltages (1.28\,MHz). We therefore re-implemented the PFB in software and added it as a new module to \vcsbeam{} (see Figure~\ref{fig:mwaxvcsbeam}). In doing this, backwards compatibility is maintained: i.e.  the PFB is only applied if the VCS observation is of the MWAX VCS type. In  the  case of legacy VCS data, the sampled voltage time series is already fine channelised (i.e. 10-kHz/100-$\mu$s resolutions).

The PFB implementation in \vcsbeam{} can be run in two additional modes that were not available in the FPGA implementations. Firstly, the output voltages can be either typecast into 4-bit complex integers (as in the FPGA implementation), or retain the full 64-bit floating point (`double') precision used during the calculation. In terms of computing resources (run-time, memory), there is no disadvantage to using the latter, since the voltages stay in memory, and are promoted to 64-bit floats during beamforming in any case. We have found, however, that the retention of the full precision makes negligible difference to the signal-to-noise ratio (\snr{}) of the final beamformed products. Secondly, the asymmetric rounding scheme of the FPGA PFB, described in the appendix of \citet{mcSweeney2020}, has been replaced with symmetric rounding; however, the original asymmetric scheme is still available as an option. To keep the data products from SMART as homogeneous as possible, we have included a specialised `SMART' option which reproduces the FPGA implementation of the PFB exactly. The deleterious effects of demoting to 4-bit integers and using an asymmetric rounding scheme are found to be marginal.


Some aspects of \vcsbeam{}  have been improved to make better use of the available compute resources, especially when executed on a supercomputing infrastructure. For instance, \vcsbeam{}  makes use of Message-Passing Interface (MPI) to handle the embarrassingly parallel processing of multiple coarse channels, as well as gathering the channels together to write out a single \psrfits{}  file. Multiple seconds of data, corresponding to several files for each coarse channel, are read in parallel using CPU threads and buffered into computer memory. The approach leverages the high data-transfer bandwidth of a distributed filesystem while limiting the impact of latency and resource contention. The core beamforming GPU kernel itself has also evolved to employ modern GPU programming techniques that make more efficient use of modern GPUs. 
Benchmarked on an AMD MI250X GPU of Pawsey's Setonix supercomputer, the execution time of the compute kernel has improved by a factor of 8 compared to the original GPU implementation \revtwo{on the same hardware}. Further, \vcsbeam{} can be compiled for both NVIDIA and AMD GPUs, \revtwo{allowing the software to run on the majority of modern supercomputing infrastructure.}

\section{Pulsar Data Analysis}\label{sec:dataanalysis}
Beamformed data from \vcsbeam{} can be written out either in \psrfits{} format \citep{psrchive} at \qty{100}{\us}/\qty{10}{\kHz} native resolutions (i.e. identical to the legacy VCS system) or as voltage time series in \vdif{} format\footnote{\url{https://vlbi.org/vlbi-standards/vdif/}}, in which case the data are stored as $24 \times \qty{1.28}{\MHz}$ coarse channels.
Our intent is to make these publicly available along with a number of data products and measurements that are science-ready.
The choice of these common data formats means the data can be readily processed using standard packages such as \presto{}, \dspsr{} and \psrchive{}.
While \dspsr{} can operate on both output formats, \presto{} can ingest only the \psrfits{} output. 
\rev{
Our detections are summarised in Table ~\ref{tab:detections}, where we also list the integration times ($T_\mathrm{obs}$). 
 
In the vast majority of cases, $T_\mathrm{obs}$ is in the 30--80\,min range; this variation is mainly due to different lengths of data that were downloaded or available at the time of processing.
In the cases where the pulsar was visible for only a fraction of the duration (e.g., due to effective on-source time less than 80\,min), we have chosen to analyse just the data segment where the zenith-normalised primary beam power towards the pulsar was greater than 20\%. As a result, for many of the detected pulsars, it is possible to obtain improved detections through longer targetted integrations, where applicable.
}
Below, we outline the main steps in the analysis of beamformed data to generate various data products of interest, such as the pulse profiles and folded archives, as well as various measurements relating to the pulsar emission and ISM propagation effects.

\subsection{Dedispersion and folding}
For the standard data analysis of non-recycled pulsars presented in this paper, we incoherently dedisperse and fold \psrfits{} files using \dspsr{} at the native frequency channel resolution (\qty{10}{\kHz}) with 10-second subintegrations and 256 phase bins.
\rev{
The folded data are stored in the \textsc{psrfits} archive format, enabling flexible data analysis with \psrchive{}.
In contrast with the non-recycled pulsar detections, the MSP detections in \citetalias{pasathree} were phase-coherently dedispersed and folded with up to 1024 phase bins.}
However, the flux densities, DMs, and RMs have been measured using the same methods for both the non-recycled pulsars and the MSPs.
These methods will be briefly summarised here for completeness.

\subsection{RFI mitigation}
The exceptionally radio-quiet nature of the Observatory site (in comparison to most other observatories) and our preferential strategy for SMART survey observations to be during nightly hours make the mitigation of radio-frequency interference (RFI) non-essential for the majority of routine pulsar data processing.
Indeed, sensitive pulsar searches will benefit from a careful assessment of the periodic RFI environment (and devising suitable steps to mitigate it), and this is part of the SMART processing plans under active development.
Typically, much of the SMART data are remarkably clear of RFI, and even extreme cases (due to either impulsive or narrowband type RFI) are limited to a small fraction ($\lesssim$10\%) of the data.
Importantly, having access to full-resolution data allows \presto{} tools (such as \rfifind{}) or \psrchive{} tools (such as \paz{}, \pazi{}, or \psrzap{}) to be effectively used to excise any RFI present in our data.
Oftentimes, this tends to be science-dependent, and so we prefer to retain that flexibility for users.
However, for folded data products, we have opted to perform automated RFI mitigation using \clfd{} \citep{Morello2019}.
In our testing, this is sufficient to remove the majority of obvious RFI.

\begin{figure*}[t]
    \centering
    \hfill
    \begin{subfigure}[b]{0.495\linewidth}
        \includegraphics[width=\linewidth]{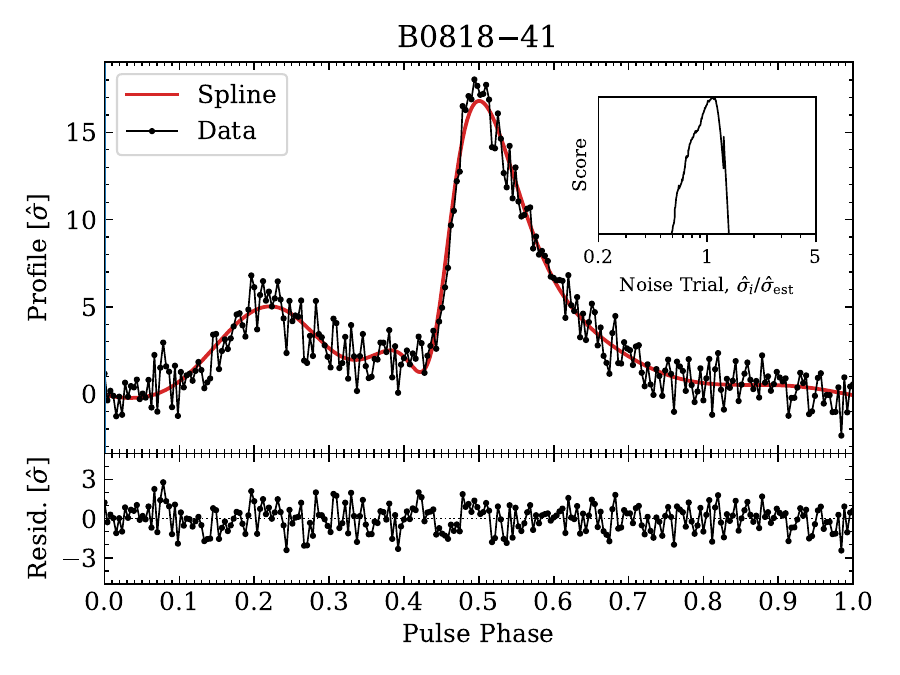}
    \end{subfigure}
    \begin{subfigure}[b]{0.495\linewidth}
        \includegraphics[width=\linewidth]{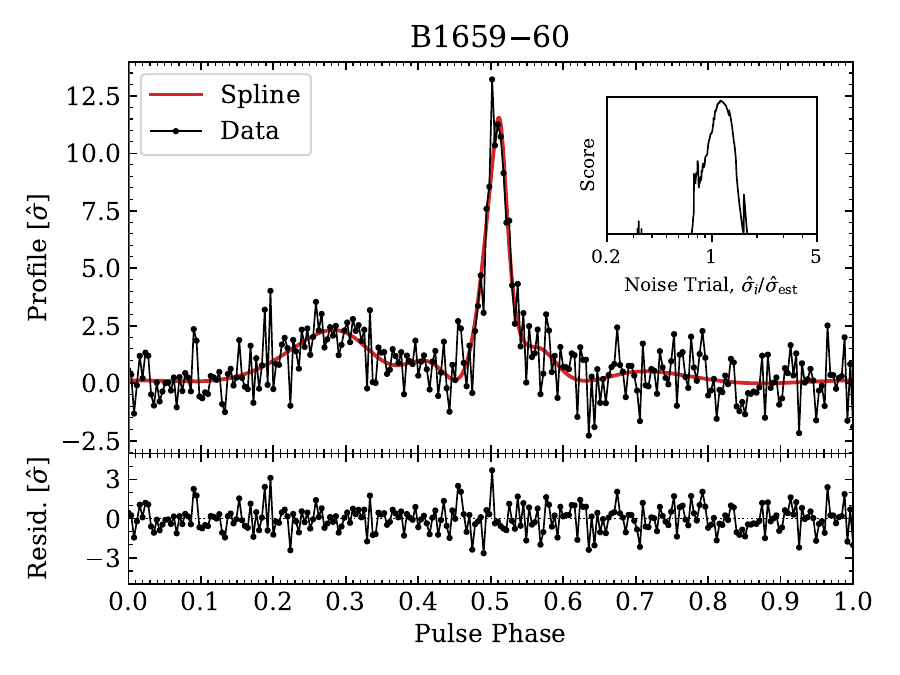}
    \end{subfigure}
    \begin{subfigure}[b]{0.495\linewidth}
        \includegraphics[width=\linewidth]{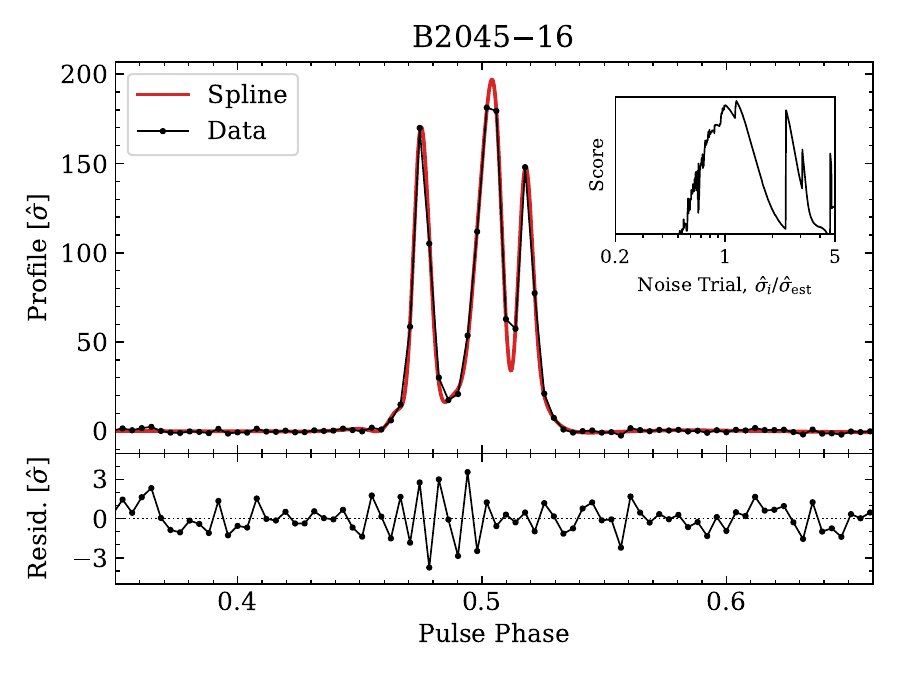}
    \end{subfigure}
    \begin{subfigure}[b]{0.495\linewidth}
        \includegraphics[width=\linewidth]{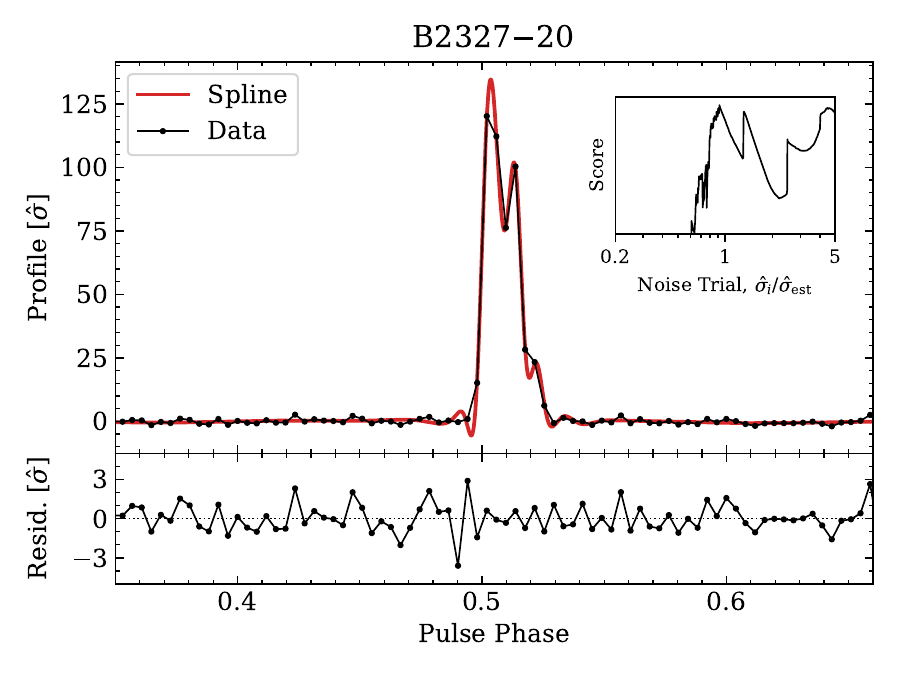}
    \end{subfigure}
    \caption{
    \rev{
    Example spline fits to the integrated pulse profiles of PSRs B0818$-$41 (top left), B1659$-$60 (top right), B2045$-$16 (bottom left), and B2327$-$20 (bottom right). The top panel of each subfigure shows the measured profile (black connected dots) and the spline fit (red lines), whilst the bottom panel shows the fit residuals. The insets show the heuristic `whiteness' score of the residuals for a range of noise trials, $\hat{\sigma}_i$, which determine the smoothing factor $s=N_\mathrm{b}\hat{\sigma}_i^2$ (see Section~\ref{sec:profiles-widths} for details). The spline fit to B2327$-$20 exhibits oscillations at the leading and trailing edges due to the sharp transitions.}
    }
    \label{fig:spline_example}
\end{figure*}

\subsection{Improved dispersion measures}\label{sec:dm}
The $\sim$20\% fractional bandwidth of SMART data (\qty{30.72}{\MHz} centred at \qty{154.24}{\MHz}) is quite comparable to previous pulsar surveys in the southern skies -- for instance, the Parkes multibeam surveys employed 288 and \qty{380}{\MHz} of bandwidth for the legacy and HTRU surveys, respectively \citep{manchester2001,keith2010}.
The inverse-frequency-square law dependence of the dispersion delay ($t _{\rm dm} \propto \nu^{-2}$, where $\nu$ is radio frequency) still means a much longer frequency lever arm in the SMART frequency band, which allows us to make precise measurements of DMs, even from a single detection with a modest \snr{}.
Indeed, higher precisions (down to $\sim 10^{-6}$\,\dmu) can be achieved when observations are made in a distributed channel set up, and with phase-coherent dedispersion techniques, as we have demonstrated in observations of MSPs \citep{kaur2019,kaur2022}.
For non-recycled pulsars in the SMART data, we typically limit our processing to incoherent dedispersion within the \qty{30.72}{\MHz} contiguous bandwidth.

The improved DMs obtainable from the SMART data are especially important for two main reasons.
Firstly, in cases where the catalogued DMs are inaccurate or imprecise, these low-frequency measurements may provide better measurements.
Secondly, in the case of frequency-dependent DMs \citep[cf.][]{Cordes2016} that arise from multipath propagation effects, DM measurements at low frequencies may differ in measurable ways to those from higher radio frequency observations \citep[e.g.][]{kaur2022}.
Aside from this subtlety, it is conceivable that profile evolution in frequency can potentially result in some systematic bias between the measured DMs.
\rev{
With observations using wide-band instrumentation becoming a common practice, this seems especially relevant \citep[e.g.][]{Hassall2012}.}
We therefore measure and tabulate DM measurements from our SMART detections.

As with our earlier partial census published in \citetalias{pasatwo},
the vast majority of our pulsar detections are in the signal-to-noise ratio (\snr{}) range $\sim$10--100 in 30--80\,min observations over the full SMART frequency band.
For the purpose of DM determination, we make use of the \pdmp{} routine from \psrchive{} to perform a fine search in period and DM.
We perform the search with 10-second subintegrations and \qty{320}{\kHz} subbands with a DM step size\footnote{Due to the nature of the \pdmp{} algorithm, the reported best DM may not be an integer multiple of the step size.} of 0.005\,\dmu{} (for non-recycled pulsars) or 0.001\,\dmu{} (for MSPs).
When a DM correction of 2$\sigma$ or greater is measured, we correct the residual dispersion delays using the \pam{} routine.
A drift in period is seldom seen in our data as we make use of most up-to-date pulsar ephemerides available, either from MeerTime\footnote{\url{https://pulsars.org.au/}} \citep{Keith2024,MTPA-zenodo} or from the ATNF Pulsar Catalogue \citep{manchester2005}.
In the rare cases where a period correction is needed, we use the \snr{}-optimised value reported by \pdmp{}.


\subsection{Integrated pulse profiles and pulse widths}\label{sec:profiles-widths}
For each pulsar detection, we average the subintegrations and frequency channels to produce an integrated pulse profile.
The detected sample of pulsars includes a diverse range of profile shapes, including a number of profiles with broadening due to multipath propagation in the ISM.
The complexity of the profiles, along with the range of \snr{} ratios, makes it challenging to reliably measure pulse widths without component fitting, which is difficult to reliably automate.
Nevertheless, measurements of pulse widths are important for studies of the intrinsic frequency-dependence of pulsar emission and scattering in the ISM.

Considering that we primarily want to model the profiles to estimate the pulse widths and onpulse phase regions, we have chosen to model each profile by fitting a smoothing spline;
\revtwo{i.e., a smooth curve constructed from multiple polynomial segments that are joined seamlessly at boundary points called `knots'.
We use the \texttt{splrep} function\footnote{\texttt{splrep} is a Python wrapper for a subroutine of the same name from the Fortran library FITPACK \citep{Dierckx1975}. \revtwo{It is documented here: \url{https://docs.scipy.org/doc/scipy/reference/generated/scipy.interpolate.splrep.html}}} from \textsc{scipy} to find the number and location of the knots, and the coefficients of the basis polynomials.
This approach enables the straight-forward computation of profile derivatives, which can be used to estimate the onpulse/offpulse.
We fit splines of degree 5 (so that the second derivative is a cubic spline) with periodic boundary conditions at the edges of the pulse profile.}

The \texttt{splrep} fitting algorithm takes a smoothing condition, $s$, which determines the `rigidity' of the spline; it should be chosen such that the noise in the profile is removed whilst retaining as much pulse structure as possible (i.e. to avoid over- or under-fitting).
In general, a good choice of smoothing condition is $s\sim N_\mathrm{b}\hat\sigma^2$, where $N_\mathrm{b}$ is the number of phase bins and $\hat\sigma$ is the standard deviation of the noise in the profile.
To find the best-fit spline, we performed a log-spaced grid search in the range $\hat\sigma \in [0.2\hat\sigma_\mathrm{est}, 5\hat\sigma_\mathrm{est}]$, where $\hat\sigma_\mathrm{est}$ is an initial estimate of $\hat\sigma$.
We determined $\hat\sigma_\mathrm{est}$ by finding the phase window of width $N_\mathrm{b}/8$ bins with the minimum integrated flux density, \rev{and measuring the standard deviation of the sample values in that window.}
For each trial $i$, we fit a spline with $s_i = N_\mathrm{b}\hat\sigma_i^2$ and performed the following tests:
\begin{itemize}
    \item a Ljung-Box test, where the null hypothesis is that all of the autocorrelations of the residuals up to a lag of 20 samples are jointly zero; and
    \item a Wald-Wolfowitz runs test, where the null hypothesis is that the sequence of signs of the residuals above or below the mean is random.
\end{itemize}
We then multiplied the $p$-values of the two tests to determine a heuristic score for the `whiteness' of the residuals\footnote{Since the two tests are not independent, the product of the $p$-values is not a valid test of the joint null hypothesis.}.
The spline with the maximum score was selected as the best fit.
\rev{Our testing indicates that this method yields reasonable fits for a wide range of profile shapes.
In Figure~\ref{fig:spline_example} we show the best-fit smoothing splines for four pulsars with a variety of profile features.
PSRs B0818$-$41 and B1659$-$60 have broad features that are well modelled by a more rigid spline, whereas PSRs B2045$-$16 and B2327$-$20 have sharper features (relative to the bin width).
The best-fit spline for B2327$-$20 exhibits `Runge's phenomenon', where sharp transitions induce oscillations in high-degree polynomials (in this case degree 5) fitted to equispaced samples.
Increasing the number of phase bins would likely improve the fits; however, we have opted to uniformly process with 256 phase bins for this work.
On the other hand, for pulsars with structure at a similar timescale to the bin width, the smoothing spline interpolates peaks and troughs in the profile (see B2045$-$16 and B2327$-$20 in Figure~\ref{fig:spline_example}).
In some cases, this may improve the estimate of the peak amplitude.
However, for the peak flux densities listed in Table~\ref{tab:detections} and Figure~\ref{fig:profile-gallery}, we have used the peak of the data rather than the spline, as we are not confident that the spline fits consistently give more accurate estimates of the peak amplitude.}

To estimate the onpulse regions, we identified the maxima of the spline which exceed $3\hat\sigma$, and then the corresponding minima of the spline which flank these maxima.
The intervals between the flanking minima were considered part of the onpulse.
If none of the samples between two of the initially identified onpulse regions fell below $1\hat\sigma$, the two regions were considered to be part of the same onpulse region.
\rev{Lastly, if multiple pulse components were identified nearby in phase (e.g. PSRs B0148$-$06 and B0957$-$47 in Figure~\ref{fig:profile-gallery}), they were considered part of the same onpulse region -- i.e. interpulses were treated separately where they could be identified, but otherwise the components were considered part of the same onpulse region.}

Spline fitting enables estimation of the profile widths by solving for the roots at a given amplitude.
As such, for all detections with peak amplitudes greater than $2\hat\sigma$ and $10\hat\sigma$, respectively, we measured $W_{50}$ and $W_{10}$ (i.e. the pulse width at 50\% and 10\% of the peak amplitude).
\rev{For consistency, the peak amplitudes used for estimating the 50\% and 10\% amplitudes were measured from the spline fits.
Each profile fit was visually inspected to verify that the pulse width measurements are reliable.}


\subsection{Mean flux densities}\label{sec:fluxdensities}
The system sensitivity was simulated following the method described by \citet{meyers2017} and \citetalias{pasathree}, which involves simulating the tied-array beam over the whole sky.
To account for the direction and frequency dependence of the beam, the sensitivity was estimated from the average of four uniformly spaced time and frequency steps, for a total of 16 simulations per observation.
It should be noted that this does not fully correct the bias introduced by beam fading (i.e. the change in on-source power over time), so we have limited our analysis to periods when the target is above 20\% beam power at 154\,MHz.
\rev{
The simulations are used to estimate the noise expected in each phase bin, $\hat{\sigma}_\mathrm{sim}$.
The mean flux density $S_\mathrm{mean}$ was then calculated as follows:
\begin{equation}
    S_\mathrm{mean} = \frac{1}{N_\mathrm{b}} \frac {\hat{\sigma}_\mathrm{sim}}{\hat{\sigma}_\mathrm{data}} \sum_{i=1}^{N_\mathrm{b}} \left( p_i - \bar{p}_\mathrm{off} \right),
\end{equation}
where $N_\mathrm{b}$ is the number of phase bins, $\hat{\sigma}_\mathrm{data}$ is the standard deviation of the noise in the profile estimated from the spline-fit residuals (see Section~\ref{sec:profiles-widths}), $p_i$ is the uncalibrated flux density of the $i$th phase bin, and $\bar{p}_\mathrm{off}$ is an estimate of the profile baseline calculated from the mean of the offpulse samples.
By estimating $\hat{\sigma}_\mathrm{data}$ from the spline-fit residuals, we are able to estimate the noise level reliably, even for pulsars with small or non-existent offpulse regions (e.g. due to scatter broadening).
For the pulsars with no offpulse samples (e.g. scattered pulsars such as B0531+21, B0833$-$45, and B1844$-$04), $\bar{p}_\mathrm{off}$ was measured from a short phase window that minimised the integrated flux density.
For each mean flux density, we assign a statistical uncertainty $\hat{\sigma}_\mathrm{sim}/\sqrt{N_\mathrm{b}}$ and a systematic uncertainty of 30\% for sources away from the Galactic plane ($|b|>\qty{10}{\degree}$) and 40\% for sources near or on the Galactic plane \citepalias[for a discussion on the sources of systematic uncertainty, see Section~3.2.1 of][]{pasathree}.
For the peak flux density, the statistical uncertainty is simply $\hat{\sigma}_\mathrm{sim}$.
The statistical and systematic uncertainties were summed in quadrature to calculate the total uncertainty.
}

\subsection{Faraday rotation measures}\label{sec:rm}
For each detection, we measured the profile-averaged RM following the method described in \citetalias{pasathree}.
To summarise, we measured the Faraday dispersion function (FDF) for all of the onpulse phase bins, then added the amplitudes of the FDFs together to make a profile-averaged FDF.
The RM was determined as the Faraday depth for which the amplitude of the FDF peaked.
This procedure was bootstrapped by adding white noise to the data to create a sample of RM point estimates.
For low-\snr{} cases with multiple peaks in the bootstrap distribution due to noise, we discarded samples outside of the RM peak, which was typically easy to identify by manually inspecting a histogram of the samples.
We then measured the mean and standard deviation of the bootstrap samples to estimate the RM and the 1$\sigma$ statistical uncertainty.

For pulsars with catalogued RMs below \qty{300}{\radian\per\m\squared}, we performed the search in the range $[-300, 300]\,\unit{\radian\per\m\squared}$ with \qty{80}{\kHz} subbands.
For pulsars with unknown RMs or catalogued RMs above \qty{300}{\radian\per\m\squared}, we performed the search in the range $[-600, 600]\,\unit{\radian\per\m\squared}$ with \qty{40}{\kHz} subbands.
The chosen frequency resolutions ensure that sensitivity is maintained within the search range \citep{Brentjens2005}.
\rev{
All searches were done with a Faraday depth resolution of \qty{0.2}{\radian\per\m\squared}, which oversamples the FDF by a factor of $\sim$12.5.

The estimated ionospheric contribution $\mathrm{RM}_{\rm iono}$ was subtracted from the observed value $\mathrm{RM}_{\rm obs}$ to estimate the RM due to the ISM, i.e. $\mathrm{RM}_{\rm ISM} = \mathrm{RM}_{\rm obs} - \mathrm{RM}_{\rm iono}$. Details on the estimation of $\mathrm{RM}_{\rm iono}$ are discussed in \citetalias{pasathree}. In short, we use \spinifex{} \citep{spinifex} for spatial and temporal interpolation of the total electron content (TEC) using a single-layer ionospheric model with global ionospheric measurements (from the JPL).
The uncertainties in $\mathrm{RM}_{\rm iono}$ calculated by \spinifex{} are doubled to better reflect the accuracy observed previously in pulsar observations at the telescope site \citep{lee2024}.
}


\begin{figure*}[p]
    \centering
    \includegraphics[width=\linewidth]{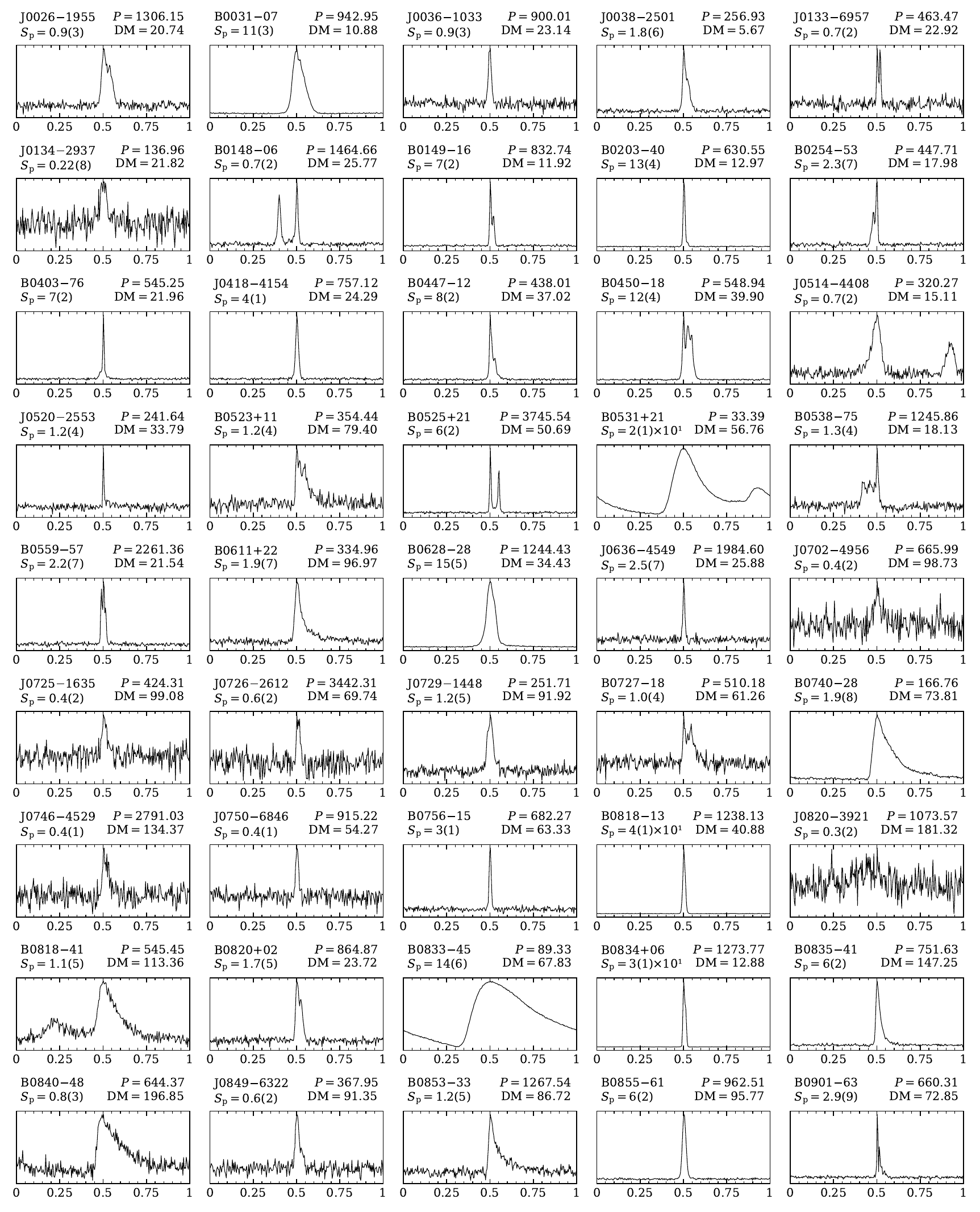}
    \caption{\rev{ Integrated pulse profiles for the best detection of each of the 205 non-recycled pulsars in SMART observations. For each pulsar, we list the B-name (or if unavailable, we use the J-name  instead), the spin period in ms, the DM in \dmu{}, and the peak flux density $S_\mathrm{p}$ in Jy (the uncertainty in the last digit is listed in parenthesis, rounded to 1 significant figure).} }
    \label{fig:profile-gallery}
\end{figure*}

\begin{figure*}[p]
    \centering
    \ContinuedFloat
    \includegraphics[width=\linewidth]{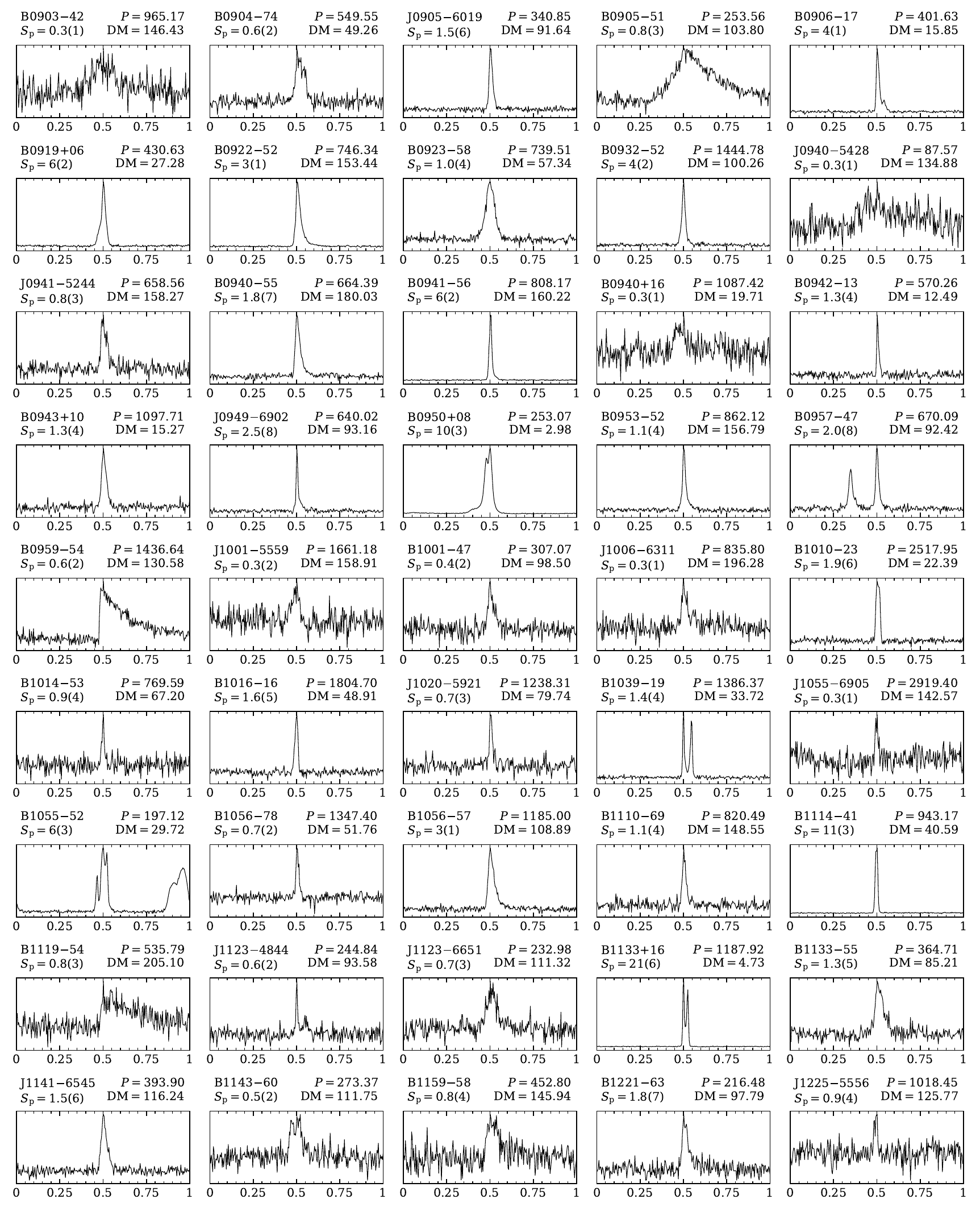}
    \caption{Continued.}
\end{figure*}

\begin{figure*}[p]
    \centering
    \ContinuedFloat
    \includegraphics[width=\linewidth]{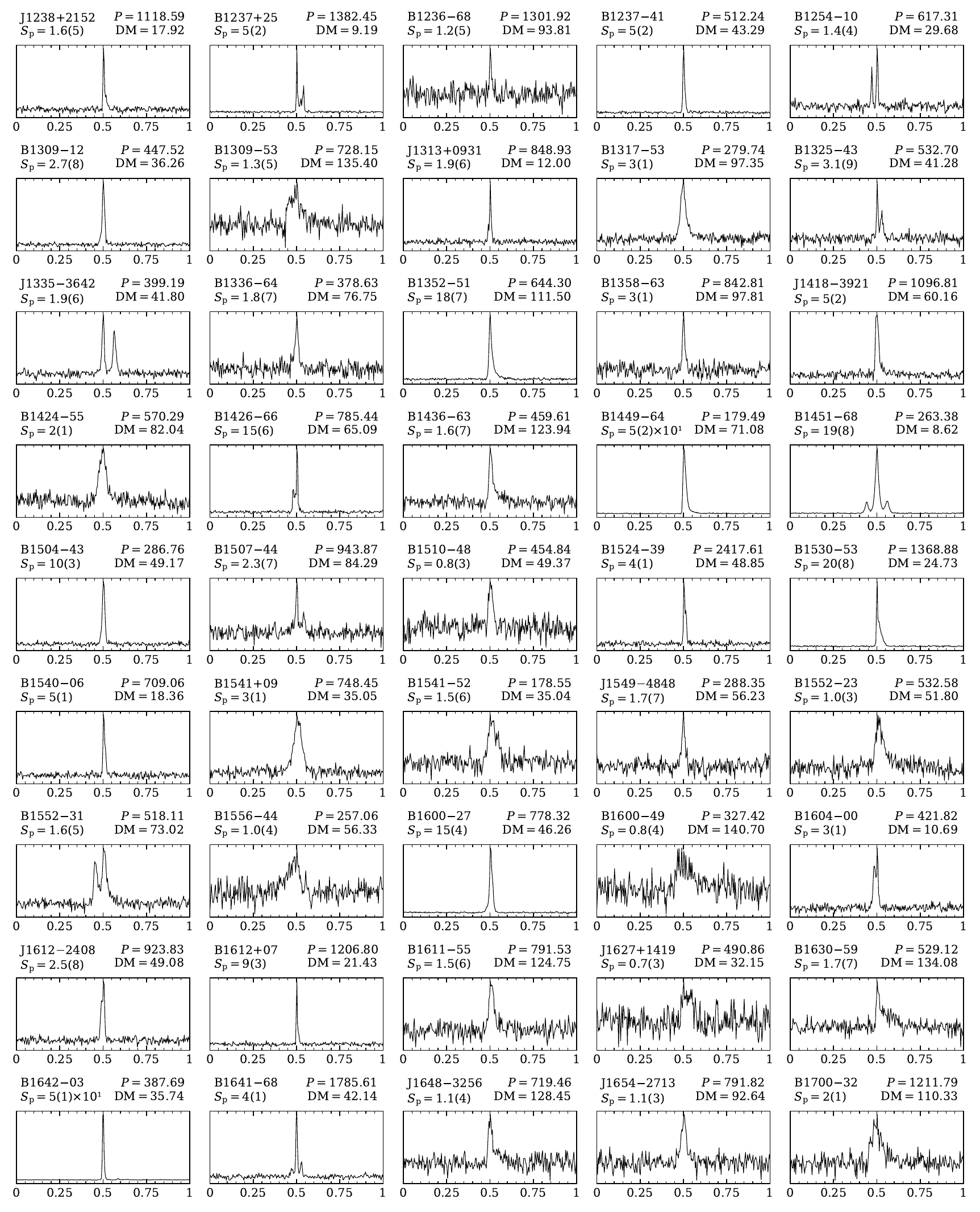}
    \caption{Continued.}
\end{figure*}

\begin{figure*}[p]
    \centering
    \ContinuedFloat
    \includegraphics[width=\linewidth]{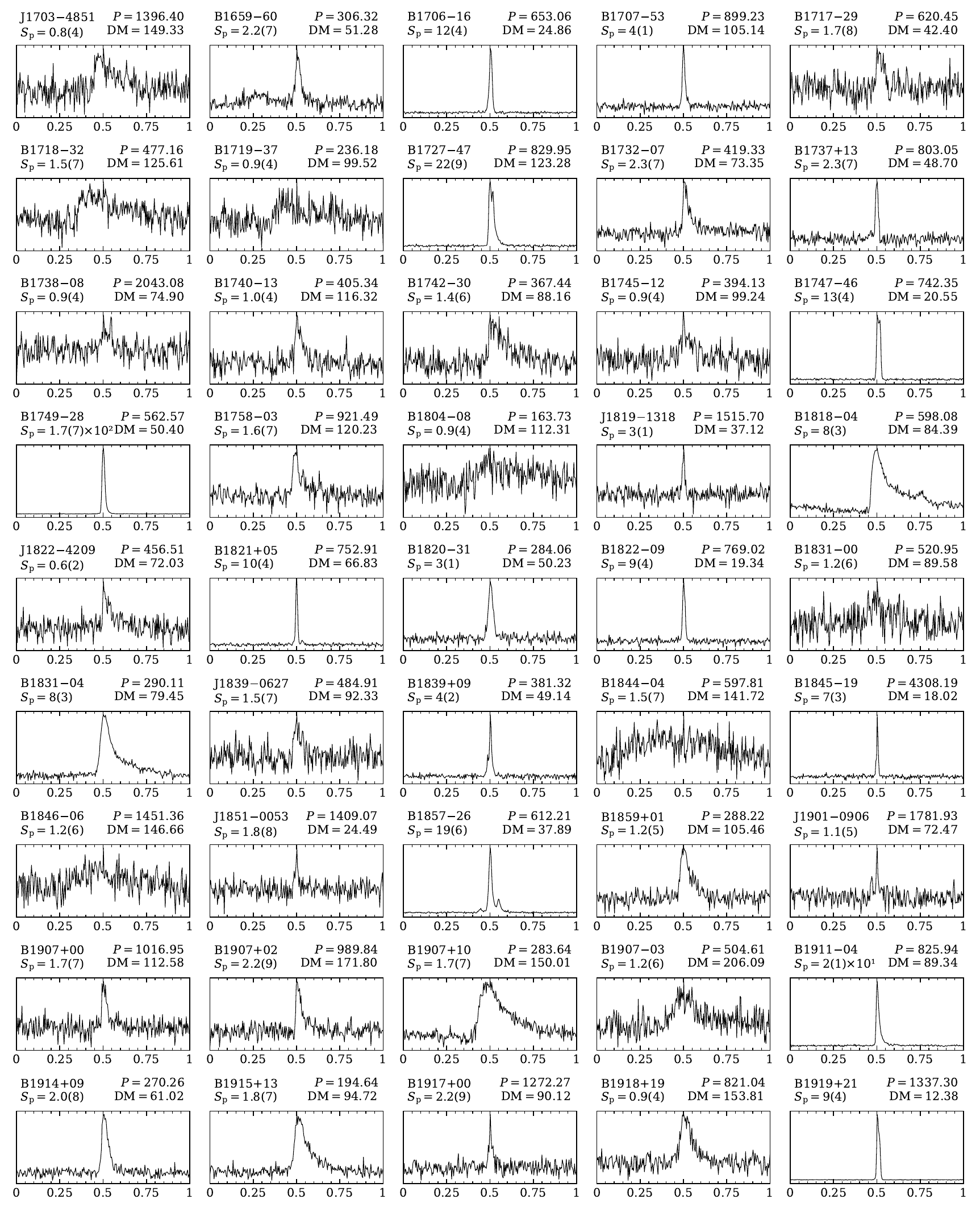}
    \caption{Continued.}
\end{figure*}

\begin{figure*}[t]
    \centering
    \ContinuedFloat
    \includegraphics[width=\linewidth]{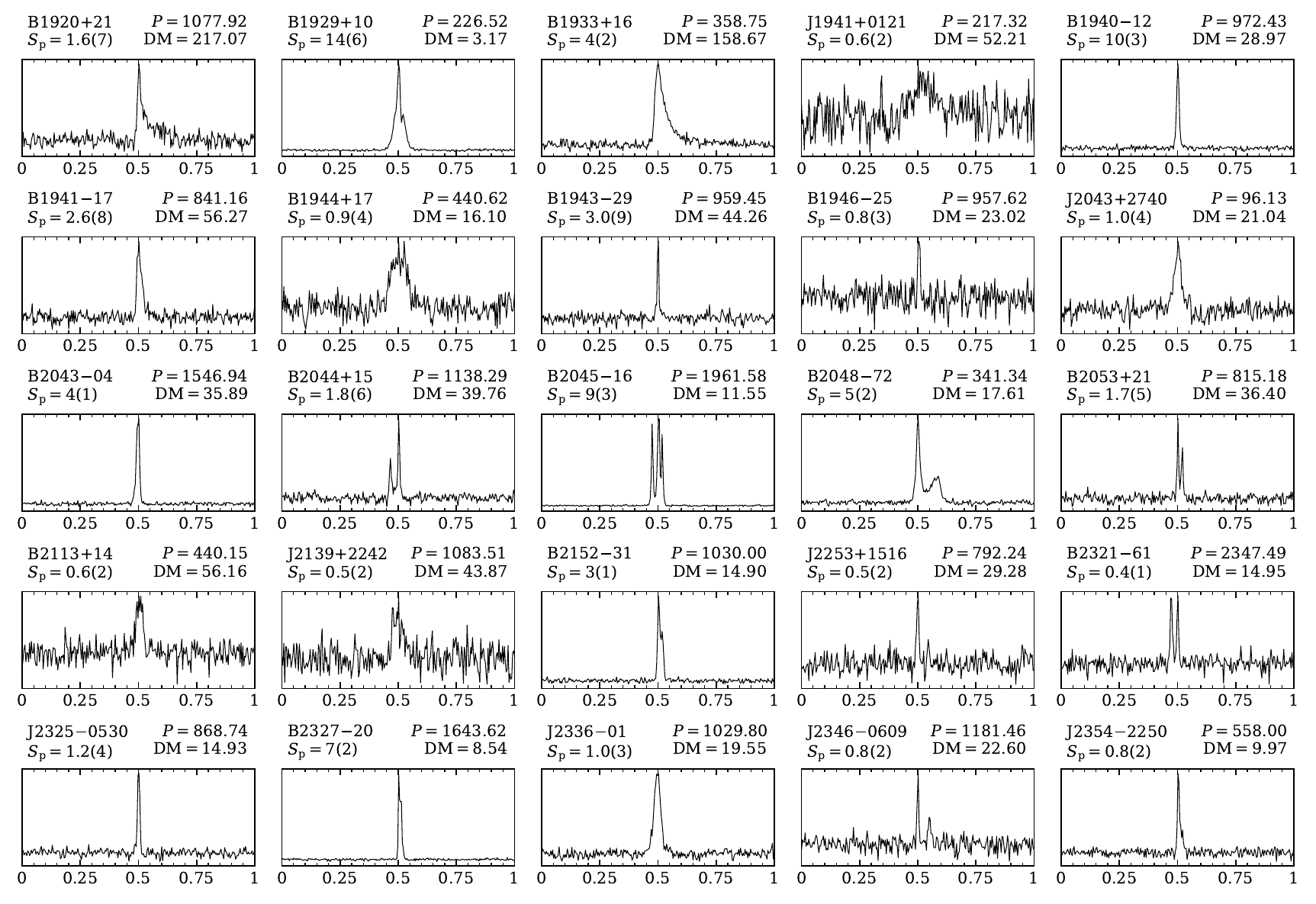}
    \caption{Continued.}
\end{figure*}


\section{Catalogue and Enhanced Data Products}\label{sec:catalogue}
Figure~\ref{fig:profile-gallery} and Table~\ref{tab:detections} provide a summary of our detections and enhanced data products derived from them. As mentioned earlier, the focus in this paper is \rev{205} known non-recycled pulsars detected in SMART data from the processing so far (two of these being earlier SMART discoveries from the shallow survey). Similar information for 40 MSPs have been presented in \citetalias{pasathree}, while new pulsars discovered in the search processing so far will be covered in a forthcoming publication. In total, we have \rev{detected 245 known pulsars} thus far in the SMART band, and counting \rev{23 additional pulsar discoveries from the SMART survey that have not been included in the census (see Section~\ref{sec:datacollection}, Figure~\ref{fig:smartprogress}), we have 268 pulsars detected in the SMART data}.
In effect, the combination of \citetalias{pasathree} and Paper~IV (this work), and the forthcoming one (on new discoveries), represent the first data release, or SMART DR1. While the processing strategies are slightly different for the two samples in Papers III and IV (e.g. coherent vs incoherent dedispersion), we maintain uniformity in the data sets and measurements presented. These are intended to be a living resource as we add more when new detections flow in from our processing.

\rev{
There have been several studies in the recent past reporting low-frequency pulsar detections using northern facilities. 
\citet{lofar2016} measured the pulse widths for a sample of 100 pulsars in the LOFAR low (15--62\,MHz) and high (120--167\,MHz) bands, and more recently, \citet{lwa2025} report a census of 100 pulsars in the LWA 20--88\,MHz band.
Low-frequency flux density measurements were reported by \citet{bilous2016} for 158 pulsars using LOFAR high band observations (110--188\,MHz), and subsequently \citep{bilous2020} extended this work to the lower band (40--78\,MHz) for a sub-sample of pulsars.
\citet{lwa2025} also report flux density measurements for their detections down to 20\,MHz.
More recently, \citet{kumar2026} extended the work of \citet{lee2022}, reporting pulse profiles and flux densities for 120 pulsars using the EDA2, an SKA-Low prototyping station comprising 256 MWA dipoles ($\sim$10\% of the Phase II MWA sensitivity for the same bandwidth).
Additionally, \citet{bsa2000} report detections and flux densities for 235 pulsars at 102.5\,MHz with the Large Phased Array (LPA or BSA).
More recently, \citet{bsa2022} report the detections of 161 northern pulsars at 111\,MHz in the Pushchino mulibeam pulsar search; however, they do not provide any profiles or flux density measurements.

The major distinction is, with the exception of EDA2 measurements, most of the previous measurements are for northern pulsars (LOFAR, LWA, or LPA/BSA) and have a wider frequency coverage (extending down to frequencies below 100\,MHz); whereas, our SMART pulsar sample is in the 140--170\,MHz band.
Detailed comparisons of overlapping sub-samples may still be useful but are outside the scope of current work.
However, we include some brief commentaries where applicable, and as such some of these aspects will be the subject of forthcoming papers. 
Together, these low-frequency samples and measurements will serve as a valuable reference for SKA-Low science verification, and more generally for exploring the frequency dependence of pulsar emission properties and/or flux-density spectra at lower frequencies.
}

Furthermore, the overlap design of SMART survey (see Figure~\ref{fig:smartprogress}) means a given sky patch is typically sampled 2--3 times, and hence multiple detections are possible. In principle, these can be co-added to produce better detections; however, the data-intensive nature of VCS processing (a typical SMART observation is $\sim$38--84\,TB) makes the full observation processing both tedious and time-consuming for our petabyte-scale data volumes. We therefore intend to periodically update our database and detections, as processing moves forward, and as and when new pulsars are detected or better detections are obtained on previously detected ones.


\subsection{Pulse profiles}
\begin{figure}
    \centering
    \includegraphics[width=\linewidth]{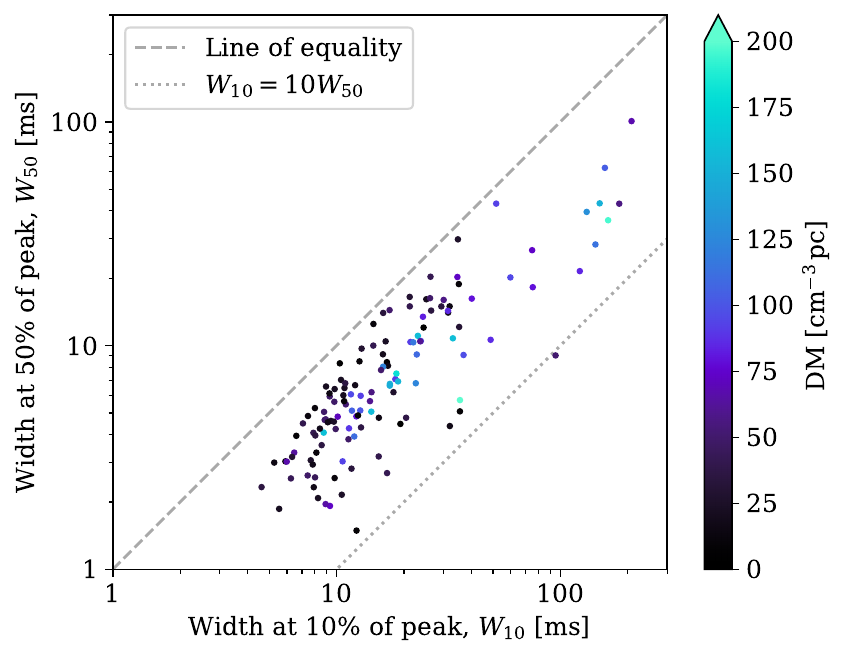}
    \caption{Scatter plot of pulse widths at 10\% ($W_{10}$) and 50\% ($W_{50}$) of the profile peak, for pulsars where both measurements could be made. All measurements come from the main pulse with the exception of PSR B1055$-$52, for which we compare both the main and interpulse. The dashed and dotted lines show where the widths are equal and where $W_{10}$ is equal to $10W_{50}$. The marker colours show the dispersion measure of each pulsar. Large deviations from equality can be due to either multipath scattering or intrinsic profile structure.}
    \label{fig:widths}
\end{figure}

The integrated pulse profiles for the best detections (by \snr{}) of the \rev{205  pulsars re-detected in the SMART data} are presented in Figure~\ref{fig:profile-gallery}.
Since the previous census reported in \citet{pasatwo}, many of the pulsar detections have been reprocessed with longer integration times and additional RFI mitigation, and hence are present here with a higher \snr{}.
Figure~\ref{fig:widths} shows the $W_{10}$--$W_{50}$ distribution for the non-recycled pulsars for which both width measurements could be made.
For two pulsars (J0514$-$4408 and J1057$-$5226), we show the measurements for both the main and interpulse.
The remaining measurements are all main pulses.
Pulsars with large differences in $W_{10}$ and $W_{50}$ generally fall into two categories: those with significant broadening due to scattering (e.g. B0853$-$33, B1818$-$04, B1920+21); and those with multiple profile components with different amplitudes (e.g. B1237+25, B1451$-$68, B1659$-$60).
The marker colours indicate the DM of each pulsar, which helps to identify which measurements fall into these categories.
Since pulsars with higher DMs are generally more scattered, we also observe a bias of higher-DM pulsars generally having larger pulse widths.

\rev{
Measurements of pulse widths at multiple frequencies can also allow the investigation of the frequency evolution of pulse profiles in attempts to  explore the emission geometries or more broadly the frequency dependence of pulsar emission. 
\citet{lofar2016} and \citet{lwa2025} present the related analysis for large pulsar samples at low frequencies with LOFAR and the LWA, spanning 15--167\,MHz and 26--88\,MHz, respectively. While both studies reveal a general widening of the pulse profiles at lower frequencies, as expected from radius-to-frequency mapping or the birefringence phenomenon, there are many exceptions, and overall there does not appear to be a specific common trend. Further, as acknowledged by \citet{lofar2016}, it is also important to deconvolve scattering tails before meaningful width comparisons are possible at low frequencies. A detailed analysis along these lines, including the comparison with the width estimates at higher frequencies, is therefore deferred to a future publication where we will report the pulse broadening analysis of our MWA sample. Nonetheless, our width estimates are robust and therefore can serve as useful indicators of profile complexity or the degree of pulse broadening, with larger values of the pulse width ratio $W_{10}/W_{50}$ implying either a multi-component pulse profile or dominant scattering tails. 
}



\begin{figure}
    \centering
    \includegraphics[width=\linewidth]{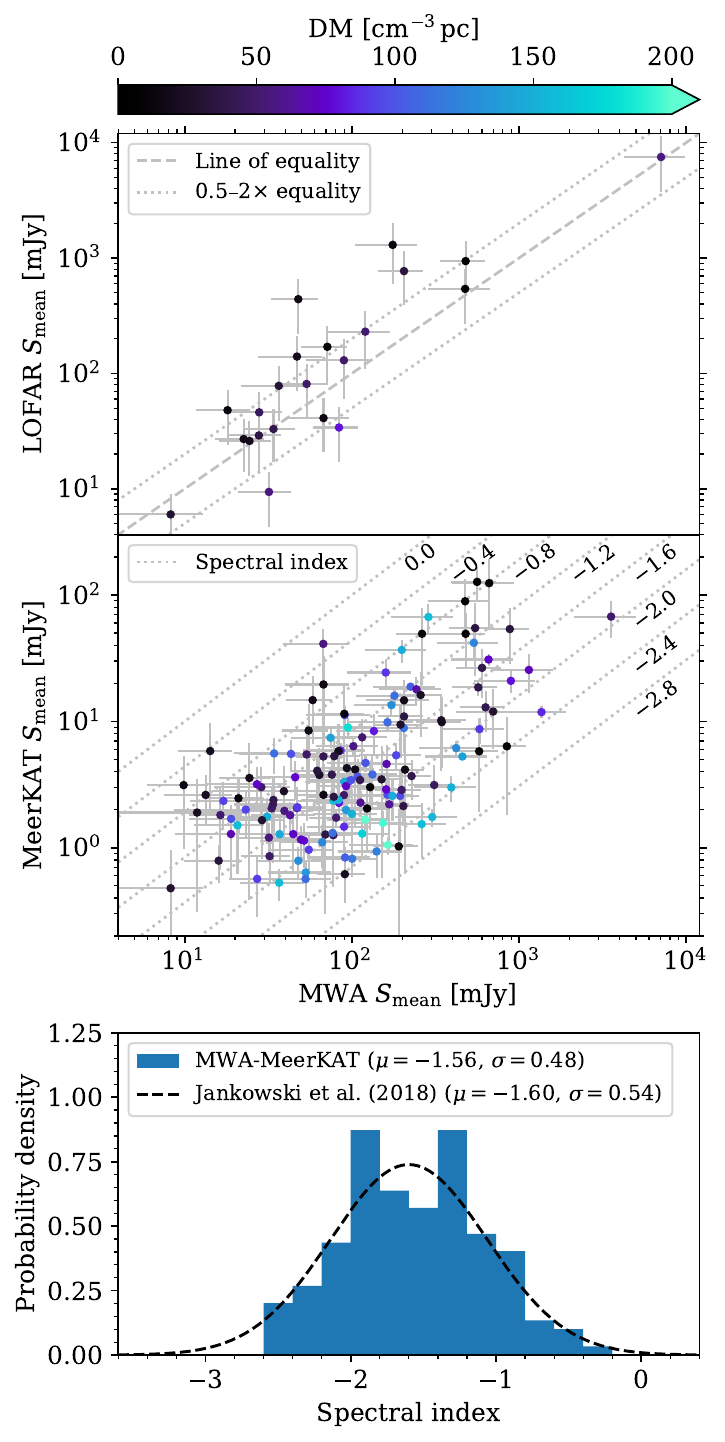}
    \caption{\rev{Comparison of the mean flux densities for non-recycled pulsars in SMART data sets (139--170\,MHz) with LOFAR mean flux densities \citep[110--188\,MHz][]{bilous2016} and long-term averaged mean flux densities from the MeerKAT Thousand-Pulsar-Array (MTPA) monitoring program \citep[896--1671\,MHz;][]{Keith2024}. 
    Top: Scatter plot of flux density measurements compared between telescopes. When comparing with LOFAR, we show the line of equality and the factor-of-two error band. When comparing with MeerKAT, we show lines indicating the two-point spectral index. The colour scale shows the dispersion measure of each pulsar. Bottom: Histogram of two-point spectral indices estimated from MWA and MeerKAT flux densities. We also show the mean spectral index distribution of the known pulsar population measured by \citet{jankowski2018}.}}
    \label{fig:flux-comparison}
\end{figure}

\subsection{Flux Densities}
Measurements of mean and peak flux densities ($S_\mathrm{mean}$ and $S_\mathrm{p}$) obtained using the method described in Section~\ref{sec:fluxdensities} are listed in Table~\ref{tab:detections}.
We emphasise that these are only first order estimates, as they do not account for any variability due to (refractive) scintillation, and moreover their estimation relies on simulating the beam patterns for telescope performance parameters.
\rev{
Nonetheless, our large sample enables us to infer broader trends in the population, particularly spectral indices and the presence of spectral flattening and turnovers.

In Figure~\ref{fig:flux-comparison}, we present a comparison of our flux densities with 22 published measurements from LOFAR \citep{bilous2016} and 149 from MeerKAT \citep{Keith2024}.
Note that the small overlapping sample with LOFAR is due to their census targeting only northern pulsars.
On average, the MWA flux densities are consistent with those from LOFAR, with the outliers likely a result of the aforementioned sources of systematic error (scintillation and calibration).
For the detections in common with MeerKAT, we measured two-point power-law spectral indices between the geometric centre frequencies\footnote{\rev{$\sqrt{\nu_\mathrm{min}\nu_\mathrm{max}}$, where $\nu_\mathrm{min}$ and $\nu_\mathrm{max}$ are the minimum and maximum frequencies of the observing band.}} of the MWA and MeerKAT  bands (154.02\,MHz and 1223.61\,MHz, respectively).
The spectral indices range from $-0.2$ to $-2.5$, with a mean of $-1.56 \pm 0.13$ (standard error) and a standard deviation $0.48$.
This is slightly shallower than the mean spectral index based on the study of a larger pulsar population, as measured by \citet{jankowski2018} to be $-1.60 \pm 0.03$ (standard error) with a standard deviation of $0.54$ when low and high frequency turnovers are modelled out.
This may be due to the lack of commonly-detected pulsars with steep spectral indices (e.g. $<-2.0$), that may be below the detectable flux density of the MeerKAT observations at $L$-band.
The lack of common steep-spectrum pulsars is also reflected in the slightly smaller standard deviation of our sample.
It is also possible that the bias is due to some pulsars turning over in the SMART frequency band.
However, even if this is the case, the observed bias is small.
This may indicate that most pulsars are turning over below 154\,MHz, or  the pulsars  exhibiting significant turnovers in the SMART frequency band were not detected.
}

\rev{
There have been several past studies that used low-frequency flux density measurements for exploring the broadband spectra of pulsars.
\citet{bilous2016} measured flux densities in the LOFAR high band for 158 pulsars and used them along with those from other published work for performing their spectral fits. LWA measurements (10--88\,MHz) have also been used for spectral analysis, though that was for modest samples \citep{stovall2015,lwa2025}. \citet{bilous2016} used single or broken power-law models for their analysis, which yielded a mean spectral index of $-1.4$. This is somewhat shallower than the mean value of $-1.56$ from our analysis and $-1.6$ from the more comprehensive analysis by \citet{jankowski2018}. 
Although the geometric centre of the LOFAR frequency band is only $\sim$10\,MHz lower than the SMART band, turnovers in the lower half of the band (i.e. between 110--150\,MHz) could significantly lower the band-averaged flux densities.
As such, the intriguing discrepancy in mean spectral index may be interpreted as an indication of spectral flattening or turnover below the SMART band. 
However, as noted by several authors \citep[e.g.][]{bilous2016,jankowski2018,bilous2020}, the quality and results from spectral fits generally depend on the number and extent (in frequency coverage) of measurements and their sparsity.
The frequency coverage of the data, along with the complexity and diversity of pulsar spectra and the uncertainties inherent to estimating low-frequency flux density measurements from beamformed observations, make the conclusions drawn from our analysis tentative.
A more detailed analysis of the spectral characteristics of our southern pulsar sample will hence be instructive, and will be reported in a future publication using the {\sc pulsar\_spectra} modelling software and catalogue \citep{pulsarspectra}.
}


\begin{figure}
    \centering
    \includegraphics[width=\linewidth]{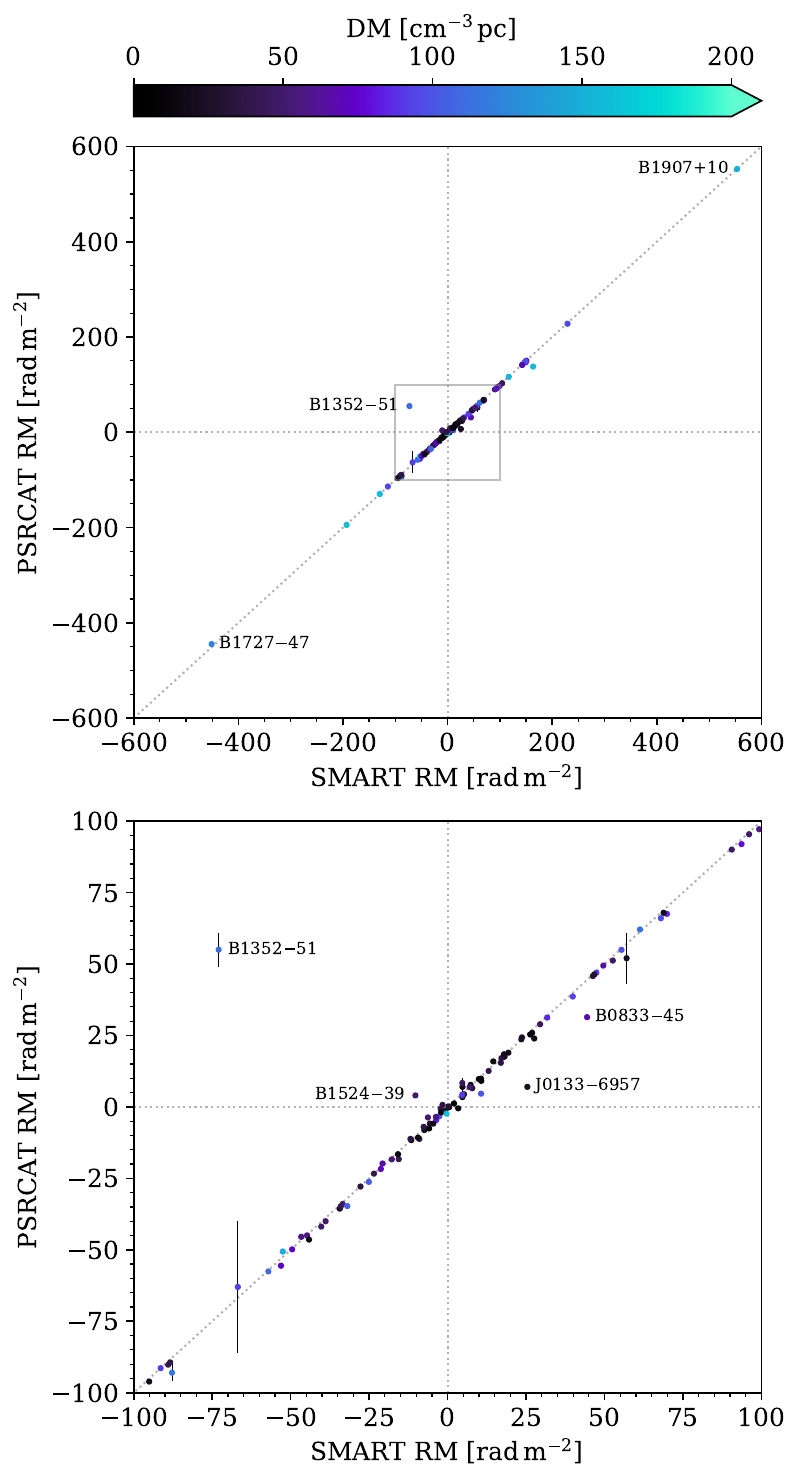}
    \caption{Comparison of RM measurements for non-recycled pulsars in SMART data sets with the RMs reported in the ATNF pulsar catalogue (i.e. PSRCAT; v2.7.0).
    The top plot shows all pulsars with both MWA and PSRCAT RMs available; the bottom plot shows just those with an RM less than \qty{100}{\rmu}, with the corresponding region indicated in the top plot.
    The highest RMs correspond to PSRs B1727$-$47 and B1907+10 (annotated), which are both located on the Galactic plane and have DMs of \qty{122.9}{\per\cm\cubed\pc} and \qty{149.8}{\per\cm\cubed\pc}, respectively.
    Several outliers are annotated: PSRs J0133$-$6957, B0833$-$45, B1352$-$51, and B1524$-$39 \citep[cf.][]{Han1999,Johnston2005,Han2006}.
    }
    \label{fig:rm-comparison}
\end{figure}

\subsection{Dispersion and Rotation Measures}
Measurements of DM and RM obtained using the methods described in Section~\ref{sec:dm} and Section~\ref{sec:rm} are listed in Table~\ref{tab:detections}. 
For the sample of non-recycled pulsars, we were able to estimate RMs for 117 pulsars. Thus, including RMs for 25 MSPs reported in \citetalias{pasathree}, we have RMs measured for 142 pulsars detected in SMART data sets.
For five pulsars (PSRs J0026$-$1955, J0038$-$2501, J0418$-$4154, J0600$-$5756, and J1612$-$2408), there are no RM measurements available in the latest pulsar catalogue \citep[v2.7.0;][]{manchester2005}; for these, we present measurements here for the first time (the locations of these pulsars are shown in Figure~\ref{fig:rm-sky}).
For the 112 non-recycled pulsars with RM measurements listed in the pulsar catalogue, Figure~\ref{fig:rm-comparison} shows a plot of our measurements against those from the pulsar catalogue \citepalias[for a similar comparison of the MSP RMs, see][]{pasathree}.

\begin{figure}[t]
    \centering
    \includegraphics[width=\linewidth]{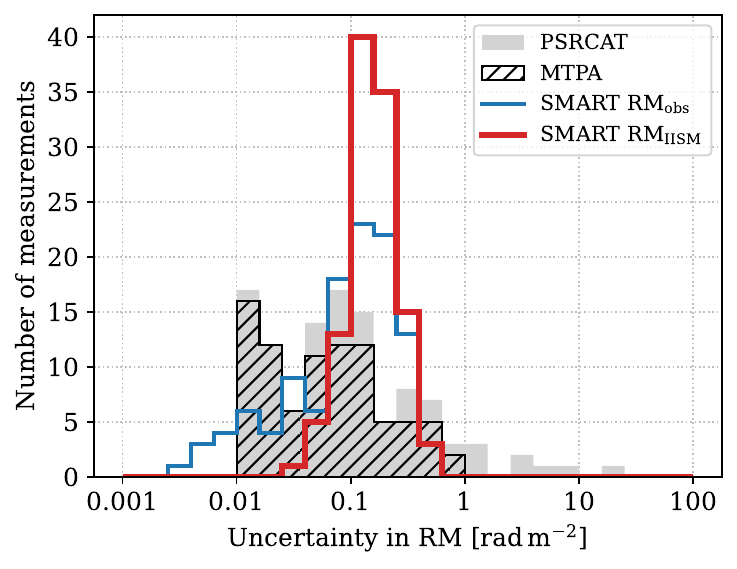}
    \caption{\rev{Histograms comparing the uncertainties in RM from this work (before and after ionospheric correction) and the ATNF pulsar catalogue (v2.7.0) for the 112 pulsars with measurements available from both. The subset of catalogue measurements from long-term monitoring with the MeerKAT Thousand Pulsar Array \citep[MTPA;][]{Keith2024} are hatched.}}
    \label{fig:rm-hist}
\end{figure}

\rev{
Low-frequency observations provide the advantage of obtaining high-precision estimates of DM and RM.
For example, Figure~\ref{fig:rm-hist} shows our RM uncertainties compared with the catalogued measurements for the same pulsars.
The majority of the catalogue measurements (86 out of 112) come from long-term monitoring with MeerKAT \citep{Keith2024}, which has enabled better precisions than can currently be obtained from single-epoch measurements due to the limited accuracy of global ionospheric models (e.g. a median uncertainty in $\mathrm{RM}_\mathrm{ISM}$ of 0.06\,\rmu{} for MeerKAT compared with 0.15\,\rmu{} for SMART).
However, as Figure~\ref{fig:rm-hist} illustrates, even without long-term monitoring, the precision that we can obtain in the SMART data is often comparable or even better than the best available measurements.
For the 29 pulsars that we have obtained more precise RM measurements than reported in the catalogue, we note this in Table~\ref{tab:detections}.
}

\begin{figure}[t]
    \centering
    \includegraphics[width=\linewidth]{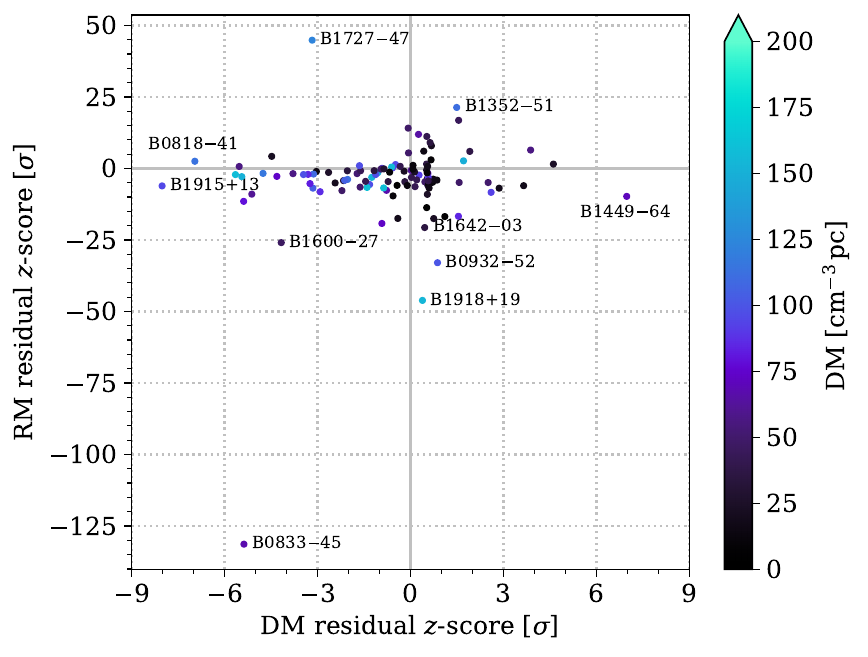}
    \caption{Distribution of residual $z$-scores (see Equation~\ref{eq:zscore}) between the RM and DM estimates from SMART data sets and from the ATNF pulsar catalogue (v2.7.0). \rev{Some significant outliers have been annotated.}}
    \label{fig:rm-dm-comparison}
\end{figure}

\begin{figure*}[t]
    \centering
    \includegraphics[width=0.9\linewidth]{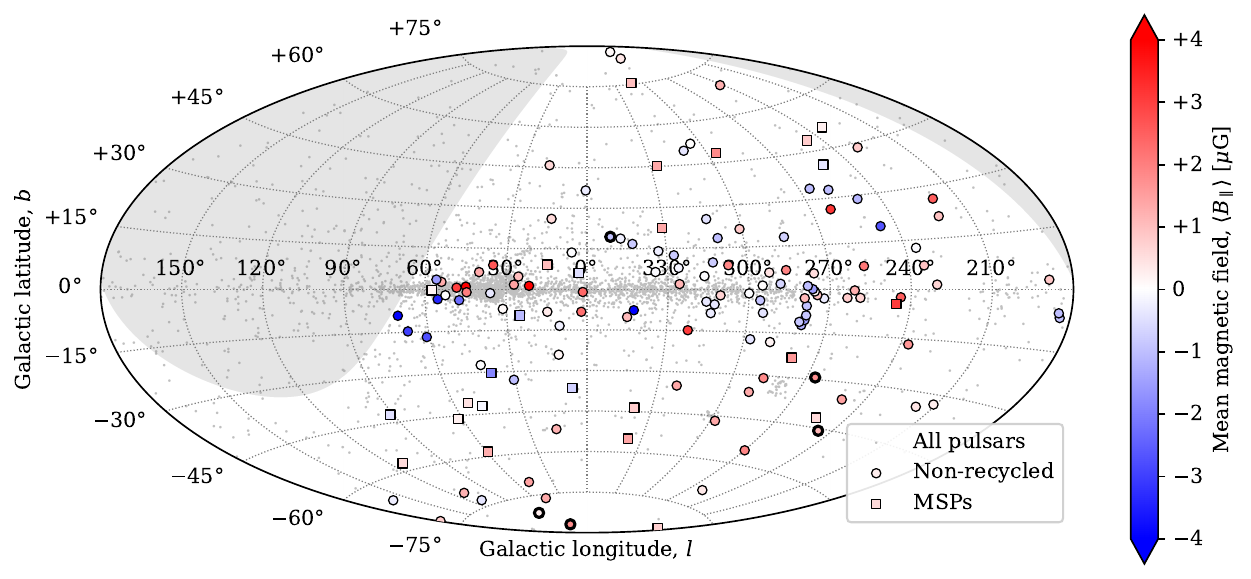}
    \caption{\rev{Galactic distribution of the line-of-sight component of the mean magnetic field $\langle B_{\parallel} \rangle$, estimated for a sample of 117 non-recycled pulsars (circles; from this work) and 25 MSPs \citepalias[squares; from][]{pasathree} using the RM and DM measurements deduced from SMART data sets. The pulsars with RMs measured for the first time in this work are emphasised.} The grey dots show all pulsars from the ATNF pulsar catalogue (v2.7.0). Declinations above $+30^\circ$, which are out of reach of the MWA, are shaded in grey.}
    \label{fig:rm-sky}
\end{figure*}

In order to compare and quantify the offsets relative to the published measurements (from the ATNF pulsar catalogue or other published work), we use the $z$-score of residuals, which is defined as
\begin{equation}\label{eq:zscore}
    z = \frac{X_\mathrm{ref} - X_\mathrm{meas}}{\sqrt{\sigma_\mathrm{ref}^2 + \sigma_\mathrm{meas}^2}},
\end{equation}
where $X_\mathrm{ref}\pm \sigma_\mathrm{ref}$ is a reference measurement (e.g. from the pulsar catalogue or published work) and $X_\mathrm{meas}\pm \sigma_\mathrm{meas}$ is the measured value from SMART data sets presented in this work. A summary of this is shown in Figure~\ref{fig:rm-dm-comparison} that shows a plot of \zdm\ vs. \zrm, where \zdm\ and \zrm\ are the $z$-scores for DM and RM, respectively. This can be used to identify the outliers. For instance, there are handful of cases with $ | \zrm |  > 20 $; e.g. for PSR B1352$-$51 (J1355$-$5153), our measured RM is \num{-72.6\pm 0.5}\,\rmu, compared with \num{55\pm 6}\,\rmu\ from \citet{Han2006}.
Our RM measurements generally show a good agreement with those from published work (see Figure~\ref{fig:rm-comparison}), and we note that a similar agreement was also noted in the earlier work of \citet{pogs2020} using image-based measurements (albeit for a much smaller sample). Given that, we are inclined to believe that there may be a sign error in the published RM for B1352$-$51.
A particularly striking case is the Vela pulsar (B0833$-$45) with $ | \zrm | > 100 $, though this is not surprising in light of recent observational evidence of the presence of small-scale magnetised plasma structure as reported in \citet{lee2024}.
These scores can be hence used to identify anomalous sight lines where either discrete and/or magnetised plasma structures are present.    
In most cases, the outliers in \zdm\ (for which we note those with $ | \zdm | > 6 $) are likely a result of inaccurate DM measurements from past observations (either due to limited S/N or small fractional bandwidths).

A summary of our RM measurements is shown in Figure~\ref{fig:rm-sky} that shows the Galactic distribution of estimated values for the mean magnetic field strength $ \langle B_{\parallel} \rangle $ towards the sight lines. The values are within $ \pm 4\, \mu$G, as we may expect for pulsars within the local ISM. A near isotropic distribution of the sample, with a sizeable fraction at high Galactic latitudes, makes them particularly useful in constraining the models for the Galactic magnetic field, both within the halo and the local ISM \rev{\citep[e.g.][]{Oswald2025}}.

\subsection{Pulse Broadening Times }
As seen from Figure~\ref{fig:profile-gallery}, many of our pulsar profiles show a significant broadening due to scattering in the ISM. 
A sizeable fraction of these is located at fairly high Galactic latitudes (see Figure~\ref{fig:smartprogress}), where there is currently a paucity of available measurements.
\rev{
These will be invaluable for constraining the model parameters towards the high-latitude regions \citep{occ20,occ2021}, which are particularly important for interpreting the measurements of fast radio bursts (FRBs). As well known, at low radio frequencies, many pulsar profiles tend to evolve to complex multi-component shapes. The deconvolution-based approaches \citep{bcc2003,young2024} are likely to be more amenable in such cases, rather than traditional forward model approach \citep{geyer2017,ooty2015}.
As the deconvolution-based methods make no assumption of the intrinsic pulse shapes, they are likely to yield more reliable estimates of pulse broadening times \citep{bhat2004}, however the peak signal-to-noise ratio needs to be sufficiently high ($\mathrm{S}/\mathrm{N} \gtrsim 20$) for them to converge. They also enable a robust estimation of the pulse broadening function (PBF), and having the PBF shapes accurately determined at multiple well-separated frequencies may also allow for the investigation of the frequency dependence of the PBF -- a subtle effect expected when inner scale effects are dominant \citep[cf.][Cordes et al. submitted]{abra2025}.} Measuring pulse broadening times and related interpretations are deferred to a future publication, where more data could be combined to increase the \snr{}.

%
\section{Discussion and Future Work}\label{sec:discussion}

With its dense coverage of the entire sky south of $+30 \degree $ in declination, substantial overlaps 
in survey pointings, and long dwell times (4800 s), SMART datasets  present an excellent resource for wide-ranging science. In particular, the novelty in data recording, where unprocessed (voltage) data are recorded (as opposed to channelised time series that has been the norm for all past and ongoing surveys) offers an enormous flexibility, with  excellent prospects  for science beyond pulsar searches. For instance, comprehensive searches can be performed for transients such as FRBs at uncommonly explored frequencies. 
A high time resolution imaging pipeline for such applications has just been developed \citep{blink2025} and is currently undergoing extensive testing and validation. The data can also be in principle processed to realise  prospective searches for signals of extra-terrestrial intelligence origin. These considerations have prompted the need to secure longer-term archiving for the foreseeable future. 

While the primary goal of SMART is to search for new pulsars and fast transients, we also process these data for re-detections of known pulsars in the 140--170\,MHz band. With the rapid progression of SKA-Low construction, a reference catalogue of low-frequency detectable pulsars would be an invaluable resource for the science commissioning and verification activities. It is also driven by some of the inherent complexities in processing VCS-recorded data given the data management and computational challenges in handling such large data rates (28--64\,\TBhr{}) and HPC requirements for post-processing. In view of the main intent of this paper, i.e. the release of a low-frequency pulsar catalogue and enhanced data products, we have limited the scope of current analysis primarily to the collation of various measurements and their broad summaries. 

Our enhanced data products include: i) pulse profiles and widths, ii) measurements of DMs and RMs, and iii) estimates of (mean and peak) flux densities. For measuring the pulse widths, we have developed a fairly robust method.  The measurements of DMs and RMs as such benefit from the strong frequency lever arm provided by our low frequencies, due to which measurement precisions are often times far superior than those obtainable through multi-epoch measurements at high frequencies using more sensitive telescopes. In general, precisions of the order of a fraction of DM unit is easily achievable for most of our detections, and sometimes down to hundredth of a DM unit, even from a single observation.  And for RMs, a measurement precision of 0.1 \rmu\ is typically achievable, though it is largely limited by our ability to account for ionospheric contribution. 

Our flux calibration methodology involves making approximations or assumptions of array beam patterns, which is unavoidable given the complexity of telescope design and array layout. Nonetheless, they are still very useful given the paucity of flux density measurements at these low frequencies for many southern pulsars. Many of our pulsar detections were made at large beam offsets (of the order of a few to several degrees from the primary beam centre; see Table~\ref{tab:detections})  due to the inherent drift-scan nature of SMART observations. Consequently, S/Ns of many of our detections are not a true reflection of the maximum sensitivity achievable with the telescope. These measurements will therefore serve as useful reference for ascertaining the scientific prospects of individual pulsars, especially in the Phase III era when longer and more sensitive pulsar observations become affordable through dedicated targeted observations.

Subsamples of our pulsars were also detected using northern telescopes; of the 205 pulsars detected in SMART data sets, 42 were detected with the LWA and 22 with LOFAR. As discussed in Section~5.2, our flux density measurements can be compared with those from LOFAR and MeerKAT \citep{bilous2016,Keith2024}. In general, DM measurements from LOFAR and LWA \citep{bilous2016,lwa2025} are significantly more precise than those obtained from our MWA detections. While the combination of a wider bandwidth  and a higher sensitivity (at $\delta > 0^{\circ}$) results in higher precisions for LOFAR DMs, the LWA measurements benefit from a larger frequency lever arm at lower frequencies (and multiple observing epochs). The possibility of frequency-dependent DMs (either due to ISM propagation effects, or unmodelled profile evolution) may also result in measurable changes (offsets) at these low frequencies. 
\revtwo{For these reasons, detailed comparisons are currently less  meaningful for the DM estimates obtained from different low-frequency instruments.}

As the processing moves forward we intend to update an electronic version of the catalogue, with the addition of more pulsars as and when they are detected in future processing. This will help to select subsamples of pulsars that are useful for broader ISM studies, besides those that are directly relevant for pulsar timing array (PTA) applications. 
\rev{ 
A substantial fraction (60\%) of listed pulsars in Table~\ref{tab:detections} are located within the local ISM ($\lesssim$3\,kpc; DMs $\lesssim$60\,\dmu); regular monitoring of these pulsars can thus facilitate a detailed characterisation and modelling of the local ISM in which most PTA pulsars are located \citep{skao-pta2025}. A multi-year time baseline of such pulsars sampling many sight lines can help to build a tomographic view of the local ISM, in an effort to improve the ISM noise modelling for PTAs, and to better quantify the ISM noise budget in timing data sets. Another aspect of direct relevance in this context is an improved modelling of the solar wind, which can benefit from regular monitoring of pulsars (both MSPs and longer period pulsars) closer to the ecliptic \citep{susarla2025}. In general, regular observations of large pulsar samples (at PTA-like cadence) will prove beneficial for probing the ISM at physical scales ($\sim$1-10\,\unit{\astronomicalunit}) that are inaccessible to other observational probes.
}

\section{Summary}\label{sec:summary}
The successful completion of data collection marks a significant milestone for the SMART project, which will emerge as the most comprehensive and sensitive pulsar survey in the southern skies in the frequency band of SKA-Low. The 4\,PB of data collected through this project essentially captures the high-time resolution sky in the pristine low-frequency band for the southern skies.
Motivated by the SKA-Low development towards science commissioning and verification, and wide-ranging science benefits provided by low-frequency measurements, we have embarked on a systematic processing of the full datasets and making the pulsar detections and enhanced data products accessible for the wider science community.  
Our processing has resulted in the detection of 245 pulsars including the 40 MSPs reported in \citetalias{pasathree}. Along with the 25 new pulsar discoveries from the searches so far, this will effectively constitute our first data release SMART DR1. 

Our measured pulse profiles in the 140--170\,MHz band are particularly valuable as the first high-fidelity records (i.e. an atlas) of a sizeable sample of southern pulsar detections at these uncommonly observed frequencies. We provide robust estimates of pulse widths at 10\% and 50\% of pulse peak (i.e. $W_{10}$ and $W_{50}$, respectively), which may serve as useful indicators of profile complexity or measurable pulse broadening. 
We also provide high-precision measurements of RMs for 142 pulsars, a significant fraction of which is located away from the plane ($|b| \gtrsim 5\deg$) and hence will be useful to constrain or improve the models for the Galactic magnetic field. Although our measured flux densities (at 154\,MHz) are only first-order estimates (as we do not account for scintillation variability),  they are still valuable for spectral analysis, considering the poorly explored pulsar spectra at these lower frequencies.   

We provide a range of data products,
including multi-channel pulse profiles, folded profiles over short sub-integrations, and measurements of DMs, RMs, and flux densities. The overlap in the SMART survey pointings is especially a major advantage, and can be leveraged for multiple detections of a pulsar, which can then be co-added for improved detections and better pulse profiles, and to obtain more reliable measurements of flux densities. In general, we have chosen the best available detection (maximum S/N) for this release, and as the processing moves forward, we intend to periodically update our data products and pulsar detections.

{%
\small

\paragraph{\bfseries Acknowledgements.}
We thank the anonymous reviewer for several useful comments that helped to improve the content and presentation of this paper.
This scientific work made use of data obtained from Inyarrimanha Ilgari Bundara, the CSIRO Murchison Radio-astronomy Observatory. We acknowledge the Wajarri Yamaji People as the Traditional Owners and Native Title Holders of the Observatory site. 
Support for the operation of the MWA is provided by the Australian Government (NCRIS), under a contract to Curtin University administered by Astronomy Australia Limited (AAL). 
CPL is supported by an Australian Government Research Training Program (RTP) Stipend and RTP Fee-Offset Scholarship (\url{https://doi.org/10.82133/C42F-K220}).
MX and QF are supported by the National SKA Program of China (2025SKA0160100).
This work was supported by resources provided by the Pawsey Supercomputing Research Centre’s Setonix Supercomputer (\url{https://doi.org/10.48569/18sb-8s43}), and their Acacia (\url{https://doi.org/10.48569/nfe9-a426}) and Banksia (\url{https://doi.org/10.48569/tnja-4s30}) Object Storage systems, with funding from the Australian Government and the Government of Western Australia.
This work was also supported by resources awarded under AAL's ASTAC merit allocation scheme on the OzSTAR national facility at the Swinburne University of Technology. 
The OzSTAR program receives funding in part from the Astronomy National Collaborative Research Infrastructure Strategy (NCRIS) allocation provided by the Australian Government.

\paragraph{\bfseries Software.}
We acknowledge the use of the following software and packages for this work:
\hyperdrive{} \citep{hyperdrive},
\casa{} \citep{casa},
\dspsr{} \citep{dspsr_ascl, dspsr},
\presto{} \citep{presto, presto_ascl},
\psrchive{} \citep{psrchive,psrchive_ascl},
\tempotwo{} \citep{tempo2_1, tempo2_ascl},
\pint{} \citep{pint_ascl, pint},
\psrpoppy{} \citep{Bates2014},
\nextflow{} \citep{nextflow},
\textsc{numpy} \citep{NumPy},
\textsc{scipy} \citep{2020SciPy-NMeth},
\textsc{astropy} \citep{Astropy2013,Astropy2018,Astropy2022},
\textsc{psrqpy} \citep{psrqpy},
\textsc{spinifex} \citep{spinifex}.

\rev{
\paragraph{\bfseries Data Availability.}
The full-polarisation folded archives for the detected pulsars are available in \textsc{psrfits} format with 256 phase bins, 30-s subintegrations, and 40-kHz frequency resolution.
The pulse profiles are also available in \textsc{ascii} format.
We have also provided the data from Tables \ref{tab:obsids} and \ref{tab:detections} in a machine-readable format.
The dataset can be accessed under the DOI: 10.5281/zenodo.20064197.
}
}

\printbibliography

\clearpage

\onecolumn

\begin{longtable}{lcccccc}
\caption{List of all 71 SMART observations. From left to right, the columns are: the SMART pointing name, the right ascension and declination of the pointing centre, the UTC date of the observation epoch, the MWA observation IDs of the VCS and calibration observations, and the number of non-flagged tiles. 
The label ID (Name) corresponds to the Blue, Red, Green, Purple and Orange pointings (see Figure~\ref{fig:smartprogress}), and SCP denotes the pointing towards the South Celestial Pole. 
} \label{tab:obsids} \\

\toprule
\multicolumn{1}{l}{Name} &
\multicolumn{1}{c}{RA} &
\multicolumn{1}{c}{Dec} &
\multicolumn{1}{c}{UTC Date} &
\multicolumn{1}{c}{VCS Obs ID} & \multicolumn{1}{c}{Cal Obs ID} &
\multicolumn{1}{c}{\#Tiles} \\
\midrule
\endfirsthead

\multicolumn{7}{c}{Continued from previous page} \\
\toprule
\multicolumn{1}{l}{Name} &
\multicolumn{1}{c}{RA} &
\multicolumn{1}{c}{Dec} &
\multicolumn{1}{c}{UTC Date} &
\multicolumn{1}{c}{VCS Obs ID} &
\multicolumn{1}{c}{Cal Obs ID} &
\multicolumn{1}{c}{\#Tiles} \\
\midrule
\endhead

\midrule
\multicolumn{7}{c}{Continued on next page} \\
\bottomrule
\endfoot

\bottomrule
\endlastfoot

B01 & $22^{\mathrm{h}}00^{\mathrm{m}}13^{\mathrm{s}}$ & $-55^\circ06{}^\prime05{}^{\prime\prime}$ & 2018-09-19 & 1221399680 & 1221405408 & 124 \\
B02 & $22^{\mathrm{h}}30^{\mathrm{m}}15^{\mathrm{s}}$ & $+01^\circ30{}^\prime28{}^{\prime\prime}$ & 2018-09-24 & 1221832280 & 1221831856 & 126 \\
B03 & $22^{\mathrm{h}}29^{\mathrm{m}}41^{\mathrm{s}}$ & $-26^\circ47{}^\prime58{}^{\prime\prime}$ & 2018-10-01 & 1222435400 & 1222435120 & 126 \\
B04 & $23^{\mathrm{h}}34^{\mathrm{m}}40^{\mathrm{s}}$ & $-13^\circ05{}^\prime29{}^{\prime\prime}$ & 2018-10-04 & 1222697776 & 1222695592 & 125 \\
B05 & $23^{\mathrm{h}}35^{\mathrm{m}}29^{\mathrm{s}}$ & $+18^\circ32{}^\prime08{}^{\prime\prime}$ & 2018-10-08 & 1223042480 & 1223038272 & 125 \\
B06 & $23^{\mathrm{h}}35^{\mathrm{m}}19^{\mathrm{s}}$ & $-40^\circ31{}^\prime14{}^{\prime\prime}$ & 2018-11-08 & 1225713560 & 1225710744 & 124 \\
B07 & $00^{\mathrm{h}}41^{\mathrm{m}}18^{\mathrm{s}}$ & $-26^\circ48{}^\prime15{}^{\prime\prime}$ & 2018-11-12 & 1226062160 & 1226055688 & 123 \\
B08 & $00^{\mathrm{h}}41^{\mathrm{m}}01^{\mathrm{s}}$ & $+01^\circ29{}^\prime50{}^{\prime\prime}$ & 2019-10-18 & 1255444104 & 1255443816 & 124 \\
B09 & $00^{\mathrm{h}}41^{\mathrm{m}}44^{\mathrm{s}}$ & $-55^\circ06{}^\prime46{}^{\prime\prime}$ & 2018-10-22 & 1224252736 & 1224277624 & 116 \\
B10 & $00^{\mathrm{h}}41^{\mathrm{m}}40^{\mathrm{s}}$ & $-72^\circ08{}^\prime41{}^{\prime\prime}$ & 2018-11-23 & 1227009976 & 1227006584 & 126 \\
B11 & $01^{\mathrm{h}}47^{\mathrm{m}}29^{\mathrm{s}}$ & $-40^\circ30{}^\prime41{}^{\prime\prime}$ & 2018-10-29 & 1224859816 & 1224874512 & 123 \\
B12 & $01^{\mathrm{h}}46^{\mathrm{m}}07^{\mathrm{s}}$ & $+18^\circ32{}^\prime49{}^{\prime\prime}$ & 2018-11-01 & 1225118240 & 1225106088 & 123 \\
B13 & $01^{\mathrm{h}}46^{\mathrm{m}}53^{\mathrm{s}}$ & $-13^\circ04{}^\prime47{}^{\prime\prime}$ & 2018-11-05 & 1225462936 & 1225450648 & 123 \\
R01 & $02^{\mathrm{h}}52^{\mathrm{m}}39^{\mathrm{s}}$ & $+01^\circ31{}^\prime31{}^{\prime\prime}$ & 2019-09-10 & 1252177744 & 1252187184 & 77 \\
R02 & $02^{\mathrm{h}}52^{\mathrm{m}}43^{\mathrm{s}}$ & $-26^\circ47{}^\prime07{}^{\prime\prime}$ & 2019-09-17 & 1252780888 & 1252780600 & 123 \\
R03 & $03^{\mathrm{h}}22^{\mathrm{m}}19^{\mathrm{s}}$ & $-55^\circ05{}^\prime06{}^{\prime\prime}$ & 2019-09-25 & 1253471952 & 1253482168 & 124 \\
R04 & $03^{\mathrm{h}}58^{\mathrm{m}}31^{\mathrm{s}}$ & $-40^\circ28{}^\prime41{}^{\prime\prime}$ & 2019-10-01 & 1253991112 & 1253990824 & 122 \\
R05 & $03^{\mathrm{h}}58^{\mathrm{m}}07^{\mathrm{s}}$ & $+18^\circ35{}^\prime07{}^{\prime\prime}$ & 2019-10-08 & 1254594264 & 1254599656 & 124 \\
R06 & $03^{\mathrm{h}}58^{\mathrm{m}}15^{\mathrm{s}}$ & $-13^\circ02{}^\prime39{}^{\prime\prime}$ & 2019-10-15 & 1255197408 & 1255202808 & 116 \\
R07 & $04^{\mathrm{h}}43^{\mathrm{m}}02^{\mathrm{s}}$ & $-72^\circ05{}^\prime12{}^{\prime\prime}$ & 2019-10-22 & 1255803168 & 1255813280 & 122 \\
R08 & $05^{\mathrm{h}}03^{\mathrm{m}}47^{\mathrm{s}}$ & $+01^\circ34{}^\prime42{}^{\prime\prime}$ & 2019-10-29 & 1256407632 & 1256417736 & 123 \\
R09 & $05^{\mathrm{h}}04^{\mathrm{m}}03^{\mathrm{s}}$ & $-26^\circ43{}^\prime54{}^{\prime\prime}$ & 2019-11-05 & 1257010784 & 1257010488 & 119 \\
R10 & $06^{\mathrm{h}}02^{\mathrm{m}}48^{\mathrm{s}}$ & $-55^\circ00{}^\prime51{}^{\prime\prime}$ & 2019-11-12 & 1257617424 & 1257626800 & 124 \\
R11 & $06^{\mathrm{h}}09^{\mathrm{m}}50^{\mathrm{s}}$ & $-40^\circ25{}^\prime02{}^{\prime\prime}$ & 2019-11-19 & 1258221008 & 1258226400 & 124 \\
R12 & $06^{\mathrm{h}}09^{\mathrm{m}}13^{\mathrm{s}}$ & $+18^\circ38{}^\prime45{}^{\prime\prime}$ & 2019-12-06 & 1259685792 & 1259691192 & 112 \\
R13 & $06^{\mathrm{h}}09^{\mathrm{m}}33^{\mathrm{s}}$ & $-12^\circ58{}^\prime59{}^{\prime\prime}$ & 2019-12-03 & 1259427304 & 1259427008 & 113 \\
G01 & $07^{\mathrm{h}}24^{\mathrm{m}}57^{\mathrm{s}}$ & $+01^\circ38{}^\prime45{}^{\prime\prime}$ & 2019-12-17 & 1260638120 & 1260637840 & 114 \\
G02 & $07^{\mathrm{h}}25^{\mathrm{m}}13^{\mathrm{s}}$ & $-26^\circ39{}^\prime49{}^{\prime\prime}$ & 2019-12-24 & 1261241272 & 1261240560 & 112 \\
G03 & $08^{\mathrm{h}}20^{\mathrm{m}}55^{\mathrm{s}}$ & $-40^\circ21{}^\prime11{}^{\prime\prime}$ & 2020-02-17 & 1265983624 & 1265983344 & 116 \\
G04 & $08^{\mathrm{h}}20^{\mathrm{m}}26^{\mathrm{s}}$ & $+18^\circ42{}^\prime20{}^{\prime\prime}$ & 2020-02-14 & 1265725128 & 1265724848 & 122 \\
G05 & $08^{\mathrm{h}}20^{\mathrm{m}}41^{\mathrm{s}}$ & $-12^\circ55{}^\prime15{}^{\prime\prime}$ & 2020-02-19 & 1266155952 & 1266155672 & 117 \\
G06 & $08^{\mathrm{h}}43^{\mathrm{m}}08^{\mathrm{s}}$ & $-54^\circ56{}^\prime15{}^{\prime\prime}$ & 2020-02-21 & 1266329600 & 1266329312 & 125 \\
G07 & $08^{\mathrm{h}}43^{\mathrm{m}}34^{\mathrm{s}}$ & $-71^\circ58{}^\prime20{}^{\prime\prime}$ & 2020-02-28 & 1266932744 & 1266938144 & 125 \\
G08 & $09^{\mathrm{h}}26^{\mathrm{m}}12^{\mathrm{s}}$ & $+01^\circ41{}^\prime38{}^{\prime\prime}$ & 2020-02-04 & 1264867416 & 1264850632 & 105 \\
G09 & $09^{\mathrm{h}}26^{\mathrm{m}}24^{\mathrm{s}}$ & $-26^\circ36{}^\prime55{}^{\prime\prime}$ & 2020-02-11 & 1265470568 & 1265475968 & 124 \\
G10 & $10^{\mathrm{h}}31^{\mathrm{m}}53^{\mathrm{s}}$ & $-40^\circ18{}^\prime57{}^{\prime\prime}$ & 2020-02-25 & 1266680784 & 1266686192 & 124 \\
G11 & $10^{\mathrm{h}}31^{\mathrm{m}}45^{\mathrm{s}}$ & $+18^\circ44{}^\prime42{}^{\prime\prime}$ & 2020-03-01 & 1267111608 & 1267111328 & 124 \\
G12 & $10^{\mathrm{h}}31^{\mathrm{m}}51^{\mathrm{s}}$ & $-12^\circ52{}^\prime57{}^{\prime\prime}$ & 2020-03-03 & 1267283936 & 1267289336 & 122 \\
G13 & $11^{\mathrm{h}}23^{\mathrm{m}}06^{\mathrm{s}}$ & $-54^\circ54{}^\prime09{}^{\prime\prime}$ & 2020-03-05 & 1267459328 & 1267459048 & 124 \\
G14 & $11^{\mathrm{h}}37^{\mathrm{m}}22^{\mathrm{s}}$ & $+01^\circ43{}^\prime04{}^{\prime\prime}$ & 2020-03-12 & 1268063336 & 1268063056 & 124 \\
G15 & $11^{\mathrm{h}}37^{\mathrm{m}}27^{\mathrm{s}}$ & $-26^\circ35{}^\prime29{}^{\prime\prime}$ & 2020-03-15 & 1268321832 & 1268327232 & 124 \\
P01 & $12^{\mathrm{h}}39^{\mathrm{m}}18^{\mathrm{s}}$ & $-12^\circ52{}^\prime13{}^{\prime\prime}$ & 2021-03-26 & 1300809400 & 1300814800 & 119 \\
P02 & $12^{\mathrm{h}}39^{\mathrm{m}}20^{\mathrm{s}}$ & $+18^\circ45{}^\prime27{}^{\prime\prime}$ & 2021-03-28 & 1300981728 & 1300981448 & 118 \\
P03 & $12^{\mathrm{h}}39^{\mathrm{m}}06^{\mathrm{s}}$ & $-71^\circ55{}^\prime55{}^{\prime\prime}$ & 2021-03-31 & 1301240224 & 1301239936 & 118 \\
P04 & $12^{\mathrm{h}}39^{\mathrm{m}}18^{\mathrm{s}}$ & $-40^\circ18{}^\prime11{}^{\prime\prime}$ & 2021-04-02 & 1301412552 & 1301412264 & 118 \\
P05 & $13^{\mathrm{h}}44^{\mathrm{m}}51^{\mathrm{s}}$ & $-26^\circ35{}^\prime50{}^{\prime\prime}$ & 2021-04-05 & 1301674968 & 1301674688 & 119 \\
P06 & $13^{\mathrm{h}}44^{\mathrm{m}}57^{\mathrm{s}}$ & $+01^\circ42{}^\prime45{}^{\prime\prime}$ & 2021-04-07 & 1301847296 & 1301847016 & 118 \\
P07 & $13^{\mathrm{h}}58^{\mathrm{m}}58^{\mathrm{s}}$ & $-54^\circ54{}^\prime37{}^{\prime\prime}$ & 2021-04-10 & 1302106648 & 1302106368 & 118 \\
P08 & $14^{\mathrm{h}}50^{\mathrm{m}}30^{\mathrm{s}}$ & $-12^\circ53{}^\prime58{}^{\prime\prime}$ & 2021-04-17 & 1302712864 & 1302712576 & 118 \\
P09 & $14^{\mathrm{h}}50^{\mathrm{m}}41^{\mathrm{s}}$ & $+18^\circ43{}^\prime45{}^{\prime\prime}$ & 2021-04-15 & 1302540536 & 1302540248 & 120 \\
P10 & $14^{\mathrm{h}}50^{\mathrm{m}}15^{\mathrm{s}}$ & $-40^\circ19{}^\prime59{}^{\prime\prime}$ & 2021-04-12 & 1302282040 & 1302281760 & 113 \\
P11 & $15^{\mathrm{h}}55^{\mathrm{m}}45^{\mathrm{s}}$ & $-26^\circ38{}^\prime02{}^{\prime\prime}$ & 2023-06-18 & 1371131072 & 1371130952 & 140 \\
P12 & $15^{\mathrm{h}}56^{\mathrm{m}}04^{\mathrm{s}}$ & $+01^\circ40{}^\prime30{}^{\prime\prime}$ & 2023-05-07 & 1367512184 & 1367512064 & 125 \\
P13 & $16^{\mathrm{h}}38^{\mathrm{m}}07^{\mathrm{s}}$ & $-72^\circ00{}^\prime10{}^{\prime\prime}$ & 2023-05-05 & 1367342464 & 1367342344 & 129 \\
P14 & $16^{\mathrm{h}}39^{\mathrm{m}}00^{\mathrm{s}}$ & $-54^\circ58{}^\prime02{}^{\prime\prime}$ & 2023-05-06 & 1367428632 & 1367428512 & 126 \\
O01 & $17^{\mathrm{h}}01^{\mathrm{m}}30^{\mathrm{s}}$ & $-12^\circ57{}^\prime08{}^{\prime\prime}$ & 2023-05-12 & 1367946928 & 1367946808 & 122 \\
O02 & $17^{\mathrm{h}}01^{\mathrm{m}}51^{\mathrm{s}}$ & $+18^\circ40{}^\prime36{}^{\prime\prime}$ & 2023-05-13 & 1368033096 & 1368032976 & 123 \\
O03 & $17^{\mathrm{h}}01^{\mathrm{m}}15^{\mathrm{s}}$ & $-40^\circ23{}^\prime08{}^{\prime\prime}$ & 2023-05-19 & 1368550080 & 1368549960 & 126 \\
O04 & $18^{\mathrm{h}}07^{\mathrm{m}}00^{\mathrm{s}}$ & $-26^\circ42{}^\prime22{}^{\prime\prime}$ & 2023-05-20 & 1368640168 & 1368640048 & 124 \\
O05 & $18^{\mathrm{h}}07^{\mathrm{m}}19^{\mathrm{s}}$ & $+01^\circ36{}^\prime15{}^{\prime\prime}$ & 2023-05-29 & 1369415648 & 1369415528 & 122 \\
O06 & $19^{\mathrm{h}}12^{\mathrm{m}}46^{\mathrm{s}}$ & $-13^\circ01{}^\prime34{}^{\prime\prime}$ & 2023-05-30 & 1369505736 & 1369510536 & 129 \\
O07 & $19^{\mathrm{h}}12^{\mathrm{m}}59^{\mathrm{s}}$ & $+18^\circ36{}^\prime12{}^{\prime\prime}$ & 2023-06-02 & 1369764224 & 1369769024 & 137 \\
O08 & $19^{\mathrm{h}}12^{\mathrm{m}}23^{\mathrm{s}}$ & $-40^\circ27{}^\prime36{}^{\prime\prime}$ & 2023-06-04 & 1369936552 & 1369941352 & 137 \\
O09 & $19^{\mathrm{h}}19^{\mathrm{m}}22^{\mathrm{s}}$ & $-55^\circ03{}^\prime26{}^{\prime\prime}$ & 2023-06-09 & 1370367808 & 1370372608 & 123 \\
O10 & $20^{\mathrm{h}}18^{\mathrm{m}}14^{\mathrm{s}}$ & $-26^\circ46{}^\prime37{}^{\prime\prime}$ & 2023-06-10 & 1370457464 & 1370457344 & 129 \\
O11 & $20^{\mathrm{h}}18^{\mathrm{m}}24^{\mathrm{s}}$ & $+01^\circ31{}^\prime56{}^{\prime\prime}$ & 2023-07-14 & 1373387040 & 1373386920 & 133 \\
O12 & $20^{\mathrm{h}}39^{\mathrm{m}}09^{\mathrm{s}}$ & $-72^\circ07{}^\prime59{}^{\prime\prime}$ & 2023-06-19 & 1371234256 & 1371234136 & 140 \\
O13 & $21^{\mathrm{h}}23^{\mathrm{m}}56^{\mathrm{s}}$ & $-13^\circ05{}^\prime17{}^{\prime\prime}$ & 2023-06-23 & 1371581520 & 1371581400 & 133 \\
O14 & $21^{\mathrm{h}}24^{\mathrm{m}}10^{\mathrm{s}}$ & $+18^\circ32{}^\prime29{}^{\prime\prime}$ & 2023-07-02 & 1372357000 & 1372361800 & 139 \\
O15 & $21^{\mathrm{h}}23^{\mathrm{m}}46^{\mathrm{s}}$ & $-40^\circ31{}^\prime20{}^{\prime\prime}$ & 2023-06-30 & 1372184672 & 1372184552 & 130 \\
SCP & $05^{\mathrm{h}}32^{\mathrm{m}}49^{\mathrm{s}}$ & $-82^\circ48{}^\prime46{}^{\prime\prime}$ & 2023-11-14 & 1384018160 & 1384018040 & 117 \\

\end{longtable}

\twocolumn

\clearpage

\onecolumn

\begin{landscape}

\begin{longtable}{llcccccccccccccc}
\caption{\rev{Summary of the 205 non-recycled pulsars re-detected in SMART observations. For each pulsar, we list the J-name, the B-name (where available), the Galactic longitude and latitude ($l$, $b$), the spin period, and the DM from \pdmp{}. For the best detection of each pulsar, we list the \snr{} reported by \psrchive{}, the mean and peak flux densities at 154\,MHz ($S_\mathrm{mean}$ and $S_\mathrm{p}$), the observed RM and the estimated (ionosphere-corrected) RM from the IISM, the estimated pulse widths at 10\% and 50\% of the peak flux density, the SMART pointing label (see Table~\ref{tab:obsids}), the integration time ($T_\mathrm{obs}$), and the mean offset from the pointing centre.}
Notes:
$^{\dagger}$ More precise than the measurement in the ATNF pulsar catalogue (v2.7.0).
$^{{\dagger\dagger}}$ No previous RM measurement available in the ATNF pulsar catalogue (v2.7.0).
$^{\ddagger}$ Main and interpulse measurements.
$^{\ddagger\ddagger}$ Since the main and interpulse overlap for the Crab pulsar, the $W_{10}$ is made across both pulses.
} \label{tab:detections} \\

\toprule
\multicolumn{1}{l}{PSR J} &
\multicolumn{1}{l}{PSR B} &
\multicolumn{1}{c}{$l$} &
\multicolumn{1}{c}{$b$} &
\multicolumn{1}{c}{Period} &
\multicolumn{1}{c}{DM} &
\multicolumn{1}{c}{S/N} &
\multicolumn{1}{c}{$S_\mathrm{mean}$} &
\multicolumn{1}{c}{$S_\mathrm{p}$} &
\multicolumn{1}{c}{$\mathrm{RM}_\mathrm{obs}$} &
\multicolumn{1}{c}{$\mathrm{RM}_\mathrm{IISM}$} &
\multicolumn{1}{c}{$W_{10}$} &
\multicolumn{1}{c}{$W_{50}$} &
\multicolumn{1}{c}{Pointing} &
\multicolumn{1}{c}{$T_\mathrm{obs}$} &
\multicolumn{1}{c}{Offset} \\
\multicolumn{1}{l}{} &
\multicolumn{1}{l}{} &
\multicolumn{1}{c}{[\unit{\degree}]} &
\multicolumn{1}{c}{[\unit{\degree}]} &
\multicolumn{1}{c}{[ms]} &
\multicolumn{1}{c}{[\unit{\per\cm\cubed\pc}]} &
\multicolumn{1}{l}{} &
\multicolumn{1}{c}{[mJy]} &
\multicolumn{1}{c}{[Jy]} &
\multicolumn{1}{c}{[\unit{\radian\per\m\squared}]} &
\multicolumn{1}{c}{[\unit{\radian\per\m\squared}]} &
\multicolumn{1}{c}{[ms]} &
\multicolumn{1}{c}{[ms]} &
\multicolumn{1}{c}{label} &
\multicolumn{1}{c}{[min]} &
\multicolumn{1}{c}{[\unit{\degree}]} \\
\midrule
\endfirsthead

\multicolumn{16}{c}{Continued from previous page} \\
\toprule
\multicolumn{1}{l}{PSR J} &
\multicolumn{1}{l}{PSR B} &
\multicolumn{1}{c}{$l$} &
\multicolumn{1}{c}{$b$} &
\multicolumn{1}{c}{Period} &
\multicolumn{1}{c}{DM} &
\multicolumn{1}{c}{S/N} &
\multicolumn{1}{c}{$S_\mathrm{mean}$} &
\multicolumn{1}{c}{$S_\mathrm{p}$} &
\multicolumn{1}{c}{$\mathrm{RM}_\mathrm{obs}$} &
\multicolumn{1}{c}{$\mathrm{RM}_\mathrm{IISM}$} &
\multicolumn{1}{c}{$W_{10}$} &
\multicolumn{1}{c}{$W_{50}$} &
\multicolumn{1}{c}{Pointing} &
\multicolumn{1}{c}{$T_\mathrm{obs}$} &
\multicolumn{1}{c}{Offset} \\
\multicolumn{1}{l}{} &
\multicolumn{1}{l}{} &
\multicolumn{1}{c}{[\unit{\degree}]} &
\multicolumn{1}{c}{[\unit{\degree}]} &
\multicolumn{1}{c}{[ms]} &
\multicolumn{1}{c}{[\unit{\per\cm\cubed\pc}]} &
\multicolumn{1}{l}{} &
\multicolumn{1}{c}{[mJy]} &
\multicolumn{1}{c}{[Jy]} &
\multicolumn{1}{c}{[\unit{\radian\per\m\squared}]} &
\multicolumn{1}{c}{[\unit{\radian\per\m\squared}]} &
\multicolumn{1}{c}{[ms]} &
\multicolumn{1}{c}{[ms]} &
\multicolumn{1}{c}{label} &
\multicolumn{1}{c}{[min]} &
\multicolumn{1}{c}{[\unit{\degree}]} \\
\midrule
\endhead

\midrule
\multicolumn{16}{c}{Continued on next page} \\
\bottomrule
\endfoot

\bottomrule
\endlastfoot

J0026$-$1955 & -- & 83.3 & $-$80.8 & 1306.15 & 20.74(11) & 44.8 & 49(15) & 0.86(26) & 3.57(10) & 4.40(15)$^{\dagger\dagger}$ & 128.4 & 82.2 & B07 & 80 & 9.0 \\
J0034$-$0721 & B0031$-$07 & 110.4 & $-$69.8 & 942.95 & 10.88(8) & 801.7 & 8.5(2.5)$\times$10$^\text{2}$ & 11.1(3.3) & 9.797(8) & 10.63(11) & 129.6 & 69.4 & B04 & 43 & 11.7 \\
J0036$-$1033 & -- & 110.0 & $-$73.0 & 900.01 & 23.14(8) & 11.6 & 26(9) & 0.91(28) & $-$8.22(11) & $-$7.41(15)$^\dagger$ & 38.1 & 22.7 & B04 & 43 & 10.9 \\
J0038$-$2501 & -- & 67.4 & $-$86.4 & 256.93 & 5.675(22) & 98.4 & 61(18) & 1.8(6) & 7.37(23) & 8.16(26)$^{\dagger\dagger}$ & 16.8 & 8.5 & B07 & 80 & 5.0 \\
J0133$-$6957 & -- & 297.7 & $-$46.7 & 463.47 & 22.92(4) & 9.6 & 18(6) & 0.70(22) & 24.94(32) & 25.40(32)$^\dagger$ & 17.8 & 11.6 & B09 & 54 & 15.6 \\
J0134$-$2937 & -- & 230.3 & $-$80.2 & 136.96 & 21.818(24) & 8.6 & 14(5) & 0.22(8) & -- & -- & -- & 11.6 & B07 & 61 & 10.1 \\
J0151$-$0635 & B0148$-$06 & 160.4 & $-$65.0 & 1464.66 & 25.83(12) & 67.3 & 64(19) & 1.8(6) & $-$6.74(5) & $-$5.64(13) & 199.4 & 170.7 & B13 & 80 & 8.5 \\
J0152$-$1637 & B0149$-$16 & 179.3 & $-$72.5 & 832.74 & 11.92(7) & 173.5 & 1.2(4)$\times$10$^\text{2}$ & 6.5(2.0) & 6.48(8) & 7.35(13)$^\dagger$ & 31.9 & 8.3 & B07 & 28 & 14.6 \\
J0206$-$4028 & B0203$-$40 & 258.6 & $-$69.6 & 630.55 & 12.97(5) & 275.2 & 1.9(6)$\times$10$^\text{2}$ & 13(4) & 2.869(32) & 3.38(8)$^\dagger$ & 14.6 & 7.5 & B11 & 30 & 1.4 \\
J0255$-$5304 & B0254$-$53 & 269.9 & $-$55.3 & 447.71 & 17.98(4) & 33.3 & 55(17) & 2.3(7) & 27.04(5) & 27.60(12)$^\dagger$ & 21.2 & 11.6 & R03 & 60 & 5.8 \\
J0401$-$7608 & B0403$-$76 & 290.3 & $-$35.9 & 545.25 & 21.96(5) & 60.1 & 93(28) & 7.4(2.2) & 23.28(10) & 23.69(12) & 17.6 & 4.4 & R07 & 30 & 4.2 \\
J0418$-$4154 & -- & 246.3 & $-$45.7 & 757.12 & 24.29(6) & 142.8 & 89(27) & 4.1(1.2) & 21.287(34) & 21.81(14)$^{\dagger\dagger}$ & 28.8 & 13.5 & R04 & 47 & 2.8 \\
J0450$-$1248 & B0447$-$12 & 211.1 & $-$32.6 & 438.01 & 37.02(4) & 167.1 & 2.0(6)$\times$10$^\text{2}$ & 8.3(2.5) & 7.24(11) & 7.87(20) & 22.0 & 7.4 & R06 & 17 & 5.0 \\
J0452$-$1759 & B0450$-$18 & 217.1 & $-$34.1 & 548.94 & 39.90(5) & 466.6 & 6.0(1.8)$\times$10$^\text{2}$ & 12(4) & 12.473(5) & 13.08(16) & 45.5 & 32.1 & R06 & 17 & 7.5 \\
J0514$-$4408 & -- & 249.5 & $-$35.4 & 320.27 & 15.207(27) & 28.7 & 76(23) & 0.81(25) & 17.26(9) & 17.76(16)$^\dagger$ & 40.0,26.7$^\ddagger$ & 18.7,4.6$^\ddagger$ & R04 & 47 & 11.7 \\
J0520$-$2553 & -- & 228.4 & $-$30.5 & 241.64 & 33.790(21) & 9.3 & 16(6) & 1.16(35) & -- & -- & 11.0 & 2.7 & R11 & 30 & 15.5 \\
J0525+1115 & B0523+11 & 192.7 & $-$13.2 & 354.44 & 79.398(30) & 29.4 & 84(26) & 1.2(4) & -- & -- & 55.6 & 22.4 & R12 & 50 & 10.4 \\
J0528+2200 & B0525+21 & 183.9 & $-$6.9 & 3745.54 & 50.69(32) & 151.9 & 1.2(5)$\times$10$^\text{2}$ & 6.0(2.4) & $-$39.769(26) & $-$38.86(18) & 252.4 & 210.8 & R12 & 50 & 7.1 \\
J0534+2200 & B0531+21 & 184.6 & $-$5.8 & 33.39 & 56.7600(29) & 470.4 & 7.1(2.8)$\times$10$^\text{3}$ & 25(10) & $-$47.538(4) & $-$46.63(17) & 23.9$^{\ddagger\ddagger}$ & 5.6 & R12 & 20 & 10.4 \\
J0536$-$7543 & B0538$-$75 & 287.2 & $-$30.8 & 1245.86 & 18.13(12) & 31.1 & 58(18) & 1.3(4) & 26.52(16) & 26.93(17) & 153.4 & 68.3 & G07 & 34 & 11.7 \\
J0600$-$5756 & B0559$-$57 & 266.5 & $-$29.3 & 2261.36 & 21.54(19) & 90.3 & 58(18) & 2.2(7) & 29.90(4) & 30.32(10)$^{\dagger\dagger}$ & 94.7 & 53.0 & R10 & 80 & 4.2 \\
J0614+2229 & B0611+22 & 188.8 & 2.4 & 334.96 & 96.975(29) & 54.6 & 1.2(5)$\times$10$^\text{2}$ & 1.9(7) & 67.022(33) & 67.94(17)$^\dagger$ & 48.3 & 11.9 & R12 & 80 & 6.4 \\
J0630$-$2834 & B0628$-$28 & 237.0 & $-$16.8 & 1244.43 & 34.44(11) & 836.8 & 8.8(2.6)$\times$10$^\text{2}$ & 14(4) & 45.653(6) & 46.28(11) & 128.6 & 69.7 & R11 & 30 & 12.4 \\
J0636$-$4549 & -- & 254.5 & $-$21.5 & 1984.60 & 25.88(17) & 11.4 & 33(12) & 2.5(7) & -- & -- & 59.5 & 23.8 & R11 & 10 & 9.3 \\
J0702$-$4956 & -- & 260.2 & $-$18.7 & 665.99 & 98.73(20) & 9.2 & 28(10) & 0.44(16) & -- & -- & -- & 56.4 & G06 & 37 & 12.9 \\
J0725$-$1635 & -- & 231.5 & $-$0.3 & 424.31 & 99.08(5) & 7.7 & 22(10) & 0.44(19) & -- & -- & -- & 22.5 & G05 & 30 & 8.3 \\
J0726$-$2612 & -- & 240.1 & $-$4.6 & 3442.31 & 69.74(29) & 7.7 & 20(10) & 0.58(25) & -- & -- & -- & 104.6 & G02 & 30 & 1.8 \\
J0729$-$1448 & -- & 230.4 & 1.4 & 251.71 & 91.920(21) & 19.2 & 55(23) & 1.2(5) & 46.69(7) & 47.42(14) & 21.0 & 10.2 & G05 & 30 & 6.7 \\
J0729$-$1836 & B0727$-$18 & 233.8 & $-$0.3 & 510.18 & 61.34(6) & 17.8 & 47(20) & 0.9(4) & -- & -- & -- & 34.4 & G02 & 30 & 8.3 \\
J0742$-$2822 & B0740$-$28 & 243.8 & $-$2.4 & 166.76 & 73.818(14) & 355.3 & 6.6(2.6)$\times$10$^\text{2}$ & 4.9(2.0) & 149.835(9) & 150.57(10) & 48.7 & 17.4 & G02 & 80 & 5.8 \\
J0746$-$4529 & -- & 259.2 & $-$10.1 & 2791.03 & 134.4(4) & 8.3 & 17(7) & 0.43(14) & -- & -- & -- & 131.4 & G06 & 68 & 12.6 \\
J0750$-$6846 & -- & 281.1 & $-$20.0 & 915.22 & 54.27(8) & 8.2 & 12(5) & 0.44(14) & -- & -- & -- & 30.0 & G07 & 80 & 5.6 \\
J0758$-$1528 & B0756$-$15 & 234.5 & 7.2 & 682.27 & 63.33(6) & 23.6 & 43(18) & 2.8(1.1) & 48.89(20) & 49.61(24) & 17.2 & 8.9 & G05 & 30 & 3.2 \\
J0820$-$1350 & B0818$-$13 & 235.9 & 12.6 & 1238.13 & 40.88(11) & 1129.4 & 6.3(1.9)$\times$10$^\text{2}$ & 37(11) & $-$2.945(19) & $-$2.21(14) & 38.1 & 19.7 & G05 & 30 & 2.1 \\
J0820$-$3921 & -- & 257.3 & $-$1.6 & 1073.57 & 181.3(8) & 9.8 & 38(16) & 0.34(16) & -- & -- & -- & -- & G02 & 30 & 14.3 \\
J0820$-$4114 & B0818$-$41 & 258.7 & $-$2.7 & 545.45 & 113.36(5) & 79.8 & 2.0(8)$\times$10$^\text{2}$ & 1.1(5) & 60.76(28) & 61.29(30) & 305.6 & 60.3 & G06 & 80 & 14.7 \\
J0823+0159 & B0820+02 & 222.0 & 21.2 & 864.87 & 23.72(7) & 32.3 & 63(19) & 1.7(5) & 17.34(22) & 18.19(27)$^\dagger$ & 56.0 & 35.3 & G05 & 30 & 15.1 \\
J0835$-$4510 & B0833$-$45 & 263.6 & $-$2.8 & 89.33 & 67.834(8) & 311.5 & 6.2(2.5)$\times$10$^\text{3}$ & 14(6) & 43.934(5) & 44.45(10) & 72.6 & 35.1 & G06 & 80 & 10.5 \\
J0837+0610 & B0834+06 & 219.7 & 26.3 & 1273.77 & 12.88(11) & 1820.1 & 5.8(1.7)$\times$10$^\text{2}$ & 33(10) & 25.574(11) & 26.33(15) & 37.1 & 24.1 & G08 & 30 & 7.6 \\
J0837$-$4135 & B0835$-$41 & 260.9 & $-$0.3 & 751.63 & 147.25(6) & 195.0 & 2.0(8)$\times$10$^\text{2}$ & 5.7(2.3) & 146.23(12) & 146.76(16) & 50.8 & 19.4 & G06 & 80 & 13.9 \\
J0842$-$4851 & B0840$-$48 & 267.2 & $-$4.1 & 644.37 & 196.85(10) & 51.8 & 1.5(6)$\times$10$^\text{2}$ & 0.75(31) & -- & -- & 411.4 & 91.4 & G06 & 80 & 7.0 \\
J0849$-$6322 & -- & 279.4 & $-$12.2 & 367.95 & 91.350(31) & 13.1 & 27(9) & 0.63(20) & $-$91.96(21) & $-$91.42(23)$^\dagger$ & 26.3 & 10.2 & G07 & 80 & 8.9 \\
J0855$-$3331 & B0853$-$33 & 256.8 & 7.5 & 1267.54 & 86.72(11) & 46.2 & 9(4)e+01 & 1.2(5) & 148.91(27) & 149.48(30) & 241.6 & 52.5 & G09 & 60 & 8.8 \\
J0856$-$6137 & B0855$-$61 & 278.6 & $-$10.4 & 962.51 & 95.76(8) & 251.2 & 1.9(6)$\times$10$^\text{2}$ & 7.2(2.2) & $-$114.65(7) & $-$114.23(11)$^\dagger$ & 43.6 & 22.8 & G06 & 80 & 7.5 \\
J0902$-$6325 & B0901$-$63 & 280.4 & $-$11.1 & 660.31 & 72.85(6) & 35.7 & 45(14) & 2.9(9) & $-$53.62(7) & $-$53.08(11) & 24.1 & 5.0 & G07 & 80 & 9.0 \\
J0904$-$4246 & B0903$-$42 & 265.1 & 2.9 & 965.17 & 146.4(4) & 16.0 & 37(15) & 0.29(13) & -- & -- & -- & 166.3 & G06 & 80 & 13.2 \\
J0904$-$7459 & B0904$-$74 & 289.7 & $-$18.3 & 549.55 & 49.26(6) & 21.2 & 33(10) & 0.59(18) & 6.56(22) & 7.00(23) & 56.2 & 35.0 & G07 & 80 & 3.7 \\
J0905$-$6019 & -- & 278.2 & $-$8.8 & 340.85 & 91.635(29) & 48.6 & 56(23) & 2.1(8) & $-$67.27(16) & $-$66.84(18)$^\dagger$ & 17.1 & 7.9 & G06 & 80 & 6.8 \\
J0907$-$5157 & B0905$-$51 & 272.2 & $-$3.0 & 253.56 & 103.80(5) & 78.0 & 2.2(9)$\times$10$^\text{2}$ & 0.82(33) & $-$25.57(14) & $-$25.09(17) & 156.6 & 61.7 & G06 & 80 & 5.4 \\
J0908$-$1739 & B0906$-$17 & 246.1 & 19.8 & 401.63 & 15.854(34) & 131.3 & 105(32) & 4.2(1.3) & $-$34.98(7) & $-$34.34(15)$^\dagger$ & 24.2 & 7.5 & G09 & 60 & 9.9 \\
J0922+0638 & B0919+06 & 225.4 & 36.4 & 430.63 & 27.28(4) & 221.0 & 2.0(6)$\times$10$^\text{2}$ & 6.2(1.8) & -- & -- & 30.2 & 10.4 & G08 & 30 & 8.8 \\
J0924$-$5302 & B0922$-$52 & 274.7 & $-$1.9 & 746.34 & 153.47(6) & 335.5 & 3.0(1.2)$\times$10$^\text{2}$ & 8.0(3.2) & 116.39(8) & 116.87(12) & 54.9 & 20.2 & G06 & 80 & 6.5 \\
J0924$-$5814 & B0923$-$58 & 278.4 & $-$5.6 & 739.51 & 57.40(6) & 68.0 & 1.0(4)$\times$10$^\text{2}$ & 1.5(6) & $-$45.209(32) & $-$44.77(9) & 86.9 & 46.2 & G06 & 80 & 6.9 \\
J0934$-$5249 & B0932$-$52 & 275.7 & $-$0.7 & 1444.78 & 100.26(12) & 58.4 & 1.1(4)$\times$10$^\text{2}$ & 3.8(1.5) & 10.19(15) & 10.67(17) & 72.0 & 29.0 & G06 & 80 & 8.0 \\
J0940$-$5428 & -- & 277.5 & $-$1.3 & 87.57 & 134.88(10) & 15.7 & 53(22) & 0.32(14) & -- & -- & -- & 36.3 & G06 & 80 & 8.4 \\
J0941$-$5244 & -- & 276.4 & 0.1 & 658.56 & 158.27(6) & 14.8 & 37(15) & 0.76(31) & $-$193.5(5) & $-$193.1(5) & 59.4 & 28.4 & G06 & 80 & 9.0 \\
J0942$-$5552 & B0940$-$55 & 278.6 & $-$2.2 & 664.39 & 180.09(6) & 71.3 & 9(4)e+01 & 2.2(9) & -- & -- & 48.0 & 19.5 & G07 & 60 & 17.0 \\
J0942$-$5657 & B0941$-$56 & 279.3 & $-$3.0 & 808.17 & 160.23(7) & 338.7 & 2.6(1.0)$\times$10$^\text{2}$ & 13(5) & 141.808(14) & 142.26(9)$^\dagger$ & 27.7 & 12.9 & G06 & 80 & 8.7 \\
J0943+1631 & B0940+16 & 216.6 & 45.4 & 1087.42 & 19.71(34) & 8.2 & 24(8) & 0.27(10) & -- & -- & -- & 110.7 & G08 & 30 & 15.0 \\
J0944$-$1354 & B0942$-$13 & 249.1 & 28.8 & 570.26 & 12.49(5) & 12.5 & 21(7) & 1.3(4) & $-$12.26(19) & $-$11.57(23) & 18.1 & 7.4 & G12 & 30 & 9.8 \\
J0946+0951 & B0943+10 & 225.4 & 43.1 & 1097.71 & 15.27(9) & 23.0 & 48(15) & 1.3(4) & -- & -- & 72.8 & 34.8 & G11 & 30 & 12.6 \\
J0949$-$6902 & -- & 287.8 & $-$11.7 & 640.02 & 93.16(5) & 56.1 & 50(15) & 2.5(8) & -- & -- & 26.6 & 7.6 & G07 & 80 & 6.3 \\
J0953+0755 & B0950+08 & 228.9 & 43.7 & 253.07 & 2.979(22) & 442.9 & 5.6(1.7)$\times$10$^\text{2}$ & 10.1(3.0) & 0.976(11) & 2.04(14) & 23.2 & 11.7 & G11 & 30 & 13.1 \\
J0955$-$5304 & B0953$-$52 & 278.3 & 1.2 & 862.12 & 156.78(7) & 41.1 & 78(31) & 2.8(1.1) & $-$130.12(19) & $-$129.65(21) & 48.1 & 17.1 & G06 & 72 & 10.2 \\
J0959$-$4809 & B0957$-$47 & 275.7 & 5.4 & 670.09 & 92.42(6) & 54.6 & 1.0(4)$\times$10$^\text{2}$ & 2.0(8) & 39.42(8) & 39.88(12) & 135.2 & 112.8 & G10 & 63 & 9.8 \\
J1001$-$5507 & B0959$-$54 & 280.2 & 0.1 & 1436.64 & 130.20(19) & 65.7 & 1.7(7)$\times$10$^\text{2}$ & 1.0(4) & -- & -- & 735.7 & 222.0 & G06 & 69 & 10.5 \\
J1001$-$5559 & -- & 280.7 & $-$0.6 & 1661.18 & 158.9(4) & 10.3 & 21(9) & 0.34(15) & -- & -- & -- & 119.8 & G07 & 42 & 17.2 \\
J1003$-$4747 & B1001$-$47 & 276.0 & 6.1 & 307.07 & 98.50(4) & 8.6 & 19(8) & 0.44(18) & -- & -- & -- & 21.6 & G10 & 63 & 9.2 \\
J1006$-$6311 & -- & 285.6 & $-$6.0 & 835.80 & 196.28(11) & 11.9 & 22(9) & 0.34(14) & -- & -- & -- & 43.4 & G07 & 80 & 11.8 \\
J1012$-$2337 & B1010$-$23 & 262.1 & 26.4 & 2517.95 & 22.39(21) & 23.1 & 55(17) & 1.9(6) & 56.41(13) & 57.04(17)$^\dagger$ & 103.0 & 69.1 & G12 & 30 & 11.3 \\
J1016$-$5345 & B1014$-$53 & 281.2 & 2.5 & 769.59 & 67.20(7) & 5.9 & 19(10) & 0.9(4) & -- & -- & -- & 26.5 & G06 & 30 & 10.0 \\
J1018$-$1642 & B1016$-$16 & 258.3 & 32.6 & 1804.70 & 48.91(15) & 13.3 & 32(10) & 1.6(5) & $-$40.93(10) & $-$40.25(15)$^\dagger$ & 68.9 & 39.5 & G12 & 30 & 4.5 \\
J1020$-$5921 & -- & 284.7 & $-$1.9 & 1238.31 & 79.58(11) & 11.3 & 35(15) & 0.9(4) & -- & -- & 59.3 & 23.3 & G06 & 54 & 12.3 \\
J1041$-$1942 & B1039$-$19 & 265.6 & 33.6 & 1386.37 & 33.71(4) & 99.6 & 62(19) & 2.6(8) & $-$28.43(6) & $-$27.75(13) & 87.5 & 75.8 & G12 & 80 & 8.9 \\
J1055$-$6905 & -- & 292.9 & $-$8.5 & 2919.40 & 142.57(26) & 12.7 & 7(4) & 0.33(14) & -- & -- & -- & 123.0 & G07 & 62 & 10.6 \\
J1057$-$5226 & B1055$-$52 & 286.0 & 6.6 & 197.12 & 29.720(17) & 146.7 & 7.0(2.8)$\times$10$^\text{2}$ & 6.3(2.5) & 46.223(11) & 46.73(10) & 16.3,27.2$^\ddagger$ & 12.7,9.3$^\ddagger$ & G13 & 54 & 3.8 \\
J1057$-$7914 & B1056$-$78 & 297.6 & $-$17.6 & 1347.40 & 51.76(11) & 7.3 & 16(6) & 0.69(21) & $-$18.17(28) & $-$17.78(29) & 57.9 & 28.7 & G07 & 80 & 10.8 \\
J1059$-$5742 & B1056$-$57 & 288.3 & 1.9 & 1185.00 & 108.89(10) & 65.1 & 1.3(5)$\times$10$^\text{2}$ & 2.8(1.1) & $-$57.58(21) & $-$57.10(23) & 105.7 & 42.2 & G13 & 54 & 3.9 \\
J1112$-$6926 & B1110$-$69 & 294.4 & $-$8.2 & 820.49 & 148.55(7) & 12.6 & 31(13) & 1.1(4) & $-$53.0(6) & $-$52.5(6) & 55.5 & 21.6 & G07 & 46 & 11.1 \\
J1116$-$4122 & B1114$-$41 & 284.5 & 18.1 & 943.17 & 40.59(8) & 385.8 & 2.3(7)$\times$10$^\text{2}$ & 11.4(3.4) & $-$23.97(7) & $-$23.47(12) & 26.1 & 16.5 & G10 & 63 & 9.2 \\
J1121$-$5444 & B1119$-$54 & 290.1 & 5.9 & 535.79 & 205.10(30) & 20.3 & 1.2(5)$\times$10$^\text{2}$ & 0.76(33) & -- & -- & -- & 130.7 & P03 & 30 & 18.1 \\
J1123$-$4844 & -- & 288.3 & 11.6 & 244.84 & 93.585(21) & 9.5 & 17(6) & 0.59(19) & -- & -- & -- & 19.2 & G10 & 59 & 12.7 \\
J1123$-$6651 & -- & 294.5 & $-$5.4 & 232.98 & 111.316(30) & 14.8 & 53(22) & 0.69(29) & -- & -- & -- & 20.6 & G07 & 23 & 12.5 \\
J1136+1551 & B1133+16 & 241.9 & 69.2 & 1187.92 & 4.73(10) & 504.2 & 4.8(1.4)$\times$10$^\text{2}$ & 21(6) & 3.569(11) & 4.75(8) & 48.0 & 38.7 & P02 & 33 & 9.9 \\
J1136$-$5525 & B1133$-$55 & 292.3 & 5.9 & 364.71 & 85.214(31) & 22.1 & 86(35) & 1.3(5) & 31.12(18) & 31.73(19) & 44.8 & 20.4 & P03 & 50 & 17.3 \\
J1141$-$6545 & -- & 295.8 & $-$3.9 & 393.90 & 116.240(34) & 20.2 & 68(28) & 1.5(6) & $-$88.31(15) & $-$87.80(16)$^\dagger$ & 33.9 & 15.9 & P03 & 80 & 8.2 \\
J1146$-$6030 & B1143$-$60 & 295.0 & 1.3 & 273.37 & 111.75(4) & 13.4 & 34(15) & 0.49(21) & 4.28(19) & 4.84(19) & -- & 26.8 & P03 & 80 & 12.7 \\
J1202$-$5820 & B1159$-$58 & 296.5 & 3.9 & 452.80 & 145.94(9) & 14.1 & 9(4)e+01 & 0.8(4) & -- & -- & -- & 50.5 & P07 & 30 & 13.0 \\
J1224$-$6407 & B1221$-$63 & 300.0 & $-$1.4 & 216.48 & 97.787(18) & 13.5 & 9(4)e+01 & 1.8(7) & $-$3.04(32) & $-$2.60(32) & -- & 9.9 & P07 & 30 & 12.7 \\
J1225$-$5556 & -- & 299.3 & 6.7 & 1018.45 & 125.77(10) & 5.7 & 34(18) & 0.9(4) & -- & -- & -- & 43.0 & P07 & 30 & 9.7 \\
J1238+2152 & -- & 272.8 & 84.0 & 1118.59 & 17.92(10) & 13.1 & 23(7) & 1.6(5) & 3.97(19) & 5.21(21) & 34.6 & 10.2 & P02 & 80 & 5.9 \\
J1239+2453 & B1237+25 & 252.5 & 86.5 & 1382.45 & 9.19(12) & 76.1 & 71(22) & 5.2(1.6) & $-$0.81(4) & 0.47(9) & 66.3 & 8.1 & P02 & 80 & 8.0 \\
J1239$-$6832 & B1236$-$68 & 301.9 & $-$5.7 & 1301.92 & 93.81(11) & 4.8 & 23(15) & 1.2(5) & -- & -- & -- & -- & P07 & 30 & 15.0 \\
J1240$-$4124 & B1237$-$41 & 300.7 & 21.4 & 512.24 & 43.29(4) & 57.5 & 77(24) & 5.4(1.6) & 28.79(15) & 29.48(15)$^\dagger$ & 12.5 & 5.1 & P04 & 80 & 4.0 \\
J1257$-$1027 & B1254$-$10 & 305.2 & 52.4 & 617.31 & 29.68(5) & 10.5 & 29(10) & 1.4(4) & $-$0.68(26) & 0.23(26) & 31.2 & 23.4 & P01 & 30 & 6.9 \\
J1311$-$1228 & B1309$-$12 & 310.7 & 50.1 & 447.52 & 36.19(4) & 36.5 & 69(21) & 3.7(1.1) & $-$12.73(9) & $-$11.84(11)$^\dagger$ & 15.0 & 6.3 & P01 & 30 & 10.0 \\
J1312$-$5402 & B1309$-$53 & 306.0 & 8.7 & 728.15 & 135.40(19) & 12.4 & 9(4)e+01 & 1.3(5) & -- & -- & -- & 58.5 & P07 & 30 & 3.3 \\
J1313+0931 & -- & 320.4 & 71.7 & 848.93 & 12.00(7) & 10.7 & 27(10) & 1.9(6) & 2.08(15) & 2.96(17) & 31.4 & 15.0 & P06 & 30 & 8.8 \\
J1320$-$5359 & B1317$-$53 & 307.3 & 8.6 & 279.74 & 97.350(24) & 26.2 & 1.6(7)$\times$10$^\text{2}$ & 3.4(1.4) & 147.96(18) & 148.48(19) & 26.0 & 11.5 & P07 & 30 & 2.2 \\
J1328$-$4357 & B1325$-$43 & 309.9 & 18.4 & 532.70 & 41.28(5) & 14.2 & 76(26) & 3.1(9) & $-$34.64(17) & $-$34.06(18) & 30.4 & 20.8 & P07 & 30 & 11.1 \\
J1335$-$3642 & -- & 312.7 & 25.3 & 399.19 & 41.805(34) & 34.2 & 70(22) & 1.9(6) & $-$16.31(15) & $-$15.57(15)$^\dagger$ & 41.0 & 31.7 & P04 & 60 & 9.9 \\
J1340$-$6456 & B1336$-$64 & 308.0 & $-$2.6 & 378.63 & 76.745(32) & 8.3 & 52(24) & 1.8(7) & -- & -- & -- & 16.2 & P07 & 30 & 10.1 \\
J1355$-$5153 & B1352$-$51 & 313.0 & 9.7 & 644.30 & 111.50(5) & 147.1 & 4.3(1.7)$\times$10$^\text{2}$ & 18(7) & $-$73.47(14) & $-$72.94(15)$^\dagger$ & 30.2 & 9.9 & P07 & 30 & 3.3 \\
J1401$-$6357 & B1358$-$63 & 310.6 & $-$2.1 & 842.81 & 97.81(7) & 10.2 & 43(22) & 2.6(1.0) & 54.94(21) & 55.40(21) & -- & 20.9 & P07 & 30 & 9.5 \\
J1418$-$3921 & -- & 320.8 & 20.5 & 1096.81 & 60.16(9) & 32.5 & 1.7(5)$\times$10$^\text{2}$ & 5.0(1.5) & -- & -- & 60.4 & 24.2 & P07 & 30 & 15.6 \\
J1428$-$5530 & B1424$-$55 & 316.4 & 4.8 & 570.29 & 82.04(5) & 20.5 & 1.4(6)$\times$10$^\text{2}$ & 2.3(1.0) & $-$4.15(18) & $-$3.63(18) & 54.2 & 30.0 & P07 & 30 & 1.4 \\
J1430$-$6623 & B1426$-$66 & 312.7 & $-$5.4 & 785.44 & 65.09(7) & 79.8 & 2.4(1.0)$\times$10$^\text{2}$ & 15(6) & $-$21.19(7) & $-$20.74(7) & 27.4 & 6.0 & P07 & 30 & 11.5 \\
J1440$-$6344 & B1436$-$63 & 314.6 & $-$3.4 & 459.61 & 123.94(4) & 11.4 & 56(24) & 1.6(7) & -- & -- & 40.6 & 12.2 & P13 & 50 & 12.5 \\
J1453$-$6413 & B1449$-$64 & 315.7 & $-$4.4 & 179.49 & 71.080(15) & 406.7 & 1.1(5)$\times$10$^\text{3}$ & 50(20) & $-$21.728(10) & $-$21.26(4) & 7.1 & 3.4 & P07 & 30 & 10.1 \\
J1456$-$6843 & B1451$-$68 & 313.9 & $-$8.5 & 263.38 & 8.620(22) & 298.5 & 6.6(2.7)$\times$10$^\text{2}$ & 19(8) & $-$2.753(26) & $-$2.05(4) & 36.6 & 5.2 & P13 & 80 & 9.1 \\
J1507$-$4352 & B1504$-$43 & 327.3 & 12.5 & 286.76 & 49.170(24) & 24.8 & 2.0(6)$\times$10$^\text{2}$ & 9.8(3.0) & $-$34.052(35) & $-$33.45(6) & 10.0 & 5.2 & P07 & 30 & 13.2 \\
J1510$-$4422 & B1507$-$44 & 327.6 & 11.7 & 943.87 & 84.29(8) & 10.4 & 70(25) & 2.3(7) & -- & -- & -- & 22.6 & P07 & 30 & 13.0 \\
J1514$-$4834 & B1510$-$48 & 325.9 & 7.8 & 454.84 & 49.37(4) & 11.9 & 29(14) & 0.78(33) & $-$6.90(25) & $-$6.27(26)$^\dagger$ & -- & 18.8 & P10 & 60 & 10.5 \\
J1527$-$3931 & B1524$-$39 & 333.0 & 14.0 & 2417.61 & 48.85(21) & 26.6 & 57(19) & 3.5(1.1) & $-$10.93(14) & $-$10.24(15)$^\dagger$ & 70.2 & 24.8 & P10 & 60 & 9.2 \\
J1534$-$5334 & B1530$-$53 & 325.7 & 1.9 & 1368.88 & 24.73(12) & 200.3 & 3.4(1.4)$\times$10$^\text{2}$ & 20(8) & 22.57(8) & 23.50(9) & 56.4 & 11.5 & P14 & 80 & 9.5 \\
J1543$-$0620 & B1540$-$06 & 0.6 & 36.6 & 709.06 & 18.36(6) & 10.3 & 83(27) & 4.8(1.5) & $-$6.10(15) & $-$4.57(17) & 21.7 & 8.1 & O01 & 23 & 13.8 \\
J1543+0929 & B1541+09 & 17.8 & 45.8 & 748.45 & 35.05(6) & 42.3 & 2.1(6)$\times$10$^\text{2}$ & 3.1(1.0) & 15.71(15) & 17.17(16) & 85.9 & 43.7 & O02 & 15 & 14.4 \\
J1544$-$5308 & B1541$-$52 & 327.3 & 1.3 & 178.55 & 35.036(24) & 16.8 & 79(33) & 1.5(6) & -- & -- & -- & 13.4 & P14 & 80 & 8.1 \\
J1549$-$4848 & -- & 330.5 & 4.3 & 288.35 & 56.227(25) & 8.4 & 39(19) & 1.7(7) & -- & -- & -- & 7.2 & P14 & 80 & 10.0 \\
J1555$-$2341 & B1552$-$23 & 348.4 & 22.5 & 532.58 & 51.80(7) & 26.4 & 81(25) & 0.97(31) & -- & -- & -- & 33.7 & P11 & 74 & 5.4 \\
J1555$-$3134 & B1552$-$31 & 342.7 & 16.8 & 518.11 & 73.02(4) & 24.8 & 92(28) & 1.6(5) & $-$50.55(11) & $-$49.51(12)$^\dagger$ & 70.2 & 41.0 & P11 & 74 & 6.7 \\
J1559$-$4438 & B1556$-$44 & 334.5 & 6.4 & 257.06 & 56.33(5) & 10.0 & 67(29) & 1.0(4) & $-$4.73(31) & $-$3.70(31) & -- & 29.0 & P14 & 80 & 12.5 \\
J1603$-$2712 & B1600$-$27 & 347.1 & 18.8 & 778.32 & 46.26(7) & 189.4 & 3.1(9)$\times$10$^\text{2}$ & 15(4) & $-$10.07(5) & $-$8.99(7) & 30.1 & 12.9 & P11 & 74 & 4.6 \\
J1604$-$4909 & B1600$-$49 & 332.2 & 2.4 & 327.42 & 140.70(12) & 11.2 & 75(32) & 0.8(4) & -- & -- & -- & 33.2 & P14 & 80 & 8.3 \\
J1607$-$0032 & B1604$-$00 & 10.7 & 35.5 & 421.82 & 10.69(4) & 18.6 & 90(29) & 3.2(1.0) & -- & -- & 20.9 & 14.0 & O01 & 32 & 14.6 \\
J1612$-$2408 & -- & 351.0 & 19.4 & 923.83 & 49.08(8) & 20.5 & 59(19) & 2.5(8) & $-$39.69(25) & $-$38.58(25)$^{\dagger\dagger}$ & 33.6 & 21.4 & P11 & 74 & 6.3 \\
J1614+0737 & B1612+07 & 20.6 & 38.2 & 1206.80 & 21.43(10) & 13.3 & 91(29) & 8.8(2.7) & -- & -- & 26.1 & 8.8 & O02 & 20 & 11.9 \\
J1615$-$5537 & B1611$-$55 & 329.0 & $-$3.5 & 791.53 & 124.75(7) & 9.6 & 66(28) & 1.5(6) & -- & -- & -- & 29.7 & P14 & 80 & 3.9 \\
J1627+1419 & -- & 30.0 & 38.3 & 490.86 & 32.15(14) & 9.5 & 37(15) & 0.69(26) & -- & -- & -- & 40.0 & O02 & 24 & 11.3 \\
J1635$-$5954 & B1630$-$59 & 327.7 & $-$8.3 & 529.12 & 134.08(14) & 12.5 & 9(4)e+01 & 1.7(7) & -- & -- & -- & 16.0 & P14 & 80 & 5.8 \\
J1645$-$0317 & B1642$-$03 & 14.1 & 26.1 & 387.69 & 35.740(33) & 481.8 & 5.5(1.6)$\times$10$^\text{2}$ & 45(14) & 15.409(33) & 16.94(7) & 7.0 & 3.5 & O01 & 78 & 11.6 \\
J1646$-$6831 & B1641$-$68 & 321.8 & $-$14.8 & 1785.61 & 42.14(15) & 19.5 & 68(21) & 3.6(1.1) & 103.65(31) & 104.38(31) & 117.3 & 18.8 & P13 & 80 & 4.0 \\
J1648$-$3256 & -- & 349.6 & 7.7 & 719.46 & 128.45(6) & 10.9 & 48(20) & 1.1(4) & -- & -- & -- & 18.9 & P11 & 53 & 11.9 \\
J1654$-$2713 & -- & 355.0 & 10.3 & 791.82 & 92.64(7) & 11.7 & 54(18) & 1.08(35) & -- & -- & -- & 31.8 & P11 & 48 & 11.0 \\
J1703$-$3241 & B1700$-$32 & 351.8 & 5.4 & 1211.79 & 110.33(18) & 19.3 & 1.6(7)$\times$10$^\text{2}$ & 2.3(1.0) & -- & -- & -- & 86.9 & O03 & 48 & 8.6 \\
J1703$-$4851 & -- & 339.0 & $-$4.5 & 1396.40 & 149.3(6) & 14.4 & 1.0(4)$\times$10$^\text{2}$ & 0.78(35) & -- & -- & -- & 206.3 & P14 & 80 & 7.9 \\
J1704$-$6016 & B1659$-$60 & 329.8 & $-$11.4 & 306.32 & 51.285(26) & 25.1 & 1.7(5)$\times$10$^\text{2}$ & 2.2(7) & -- & -- & 113.9 & 10.8 & P14 & 80 & 6.9 \\
J1709$-$1640 & B1706$-$16 & 5.8 & 13.7 & 653.06 & 24.86(6) & 172.2 & 2.6(8)$\times$10$^\text{2}$ & 12(4) & $-$2.220(25) & $-$0.85(6) & 24.1 & 11.8 & O01 & 80 & 6.6 \\
J1711$-$5350 & B1707$-$53 & 335.7 & $-$8.5 & 899.23 & 105.14(8) & 13.2 & 9(4)e+01 & 3.7(1.5) & -- & -- & 41.1 & 18.0 & P14 & 80 & 5.2 \\
J1720$-$2933 & B1717$-$29 & 356.5 & 4.2 & 620.45 & 42.40(12) & 8.1 & 1.1(5)$\times$10$^\text{2}$ & 1.7(8) & -- & -- & -- & 38.9 & O03 & 20 & 14.8 \\
J1722$-$3207 & B1718$-$32 & 354.6 & 2.5 & 477.16 & 125.61(20) & 20.9 & 4.2(1.7)$\times$10$^\text{2}$ & 1.5(7) & -- & -- & -- & 186.7 & O03 & 48 & 8.8 \\
J1722$-$3712 & B1719$-$37 & 350.5 & $-$0.5 & 236.18 & 99.52(35) & 13.2 & 1.2(5)$\times$10$^\text{2}$ & 0.9(4) & -- & -- & -- & -- & O03 & 80 & 6.4 \\
J1731$-$4744 & B1727$-$47 & 342.6 & $-$7.7 & 829.95 & 123.28(7) & 197.1 & 5.4(2.1)$\times$10$^\text{2}$ & 15(6) & $-$451.96(14) & $-$450.96(14) & 52.3 & 26.1 & P14 & 79 & 11.2 \\
J1735$-$0724 & B1732$-$07 & 17.3 & 13.3 & 419.33 & 73.35(4) & 19.2 & 1.6(5)$\times$10$^\text{2}$ & 2.3(7) & -- & -- & -- & 15.6 & O05 & 30 & 9.4 \\
J1740+1311 & B1737+13 & 37.1 & 21.7 & 803.05 & 48.70(7) & 5.7 & 54(18) & 2.3(7) & -- & -- & 27.6 & 15.9 & O02 & 30 & 8.5 \\
J1741$-$0840 & B1738$-$08 & 17.0 & 11.3 & 2043.08 & 74.9(7) & 8.7 & 46(19) & 0.93(35) & -- & -- & -- & 150.9 & O05 & 30 & 10.5 \\
J1743$-$1351 & B1740$-$13 & 12.7 & 8.2 & 405.34 & 116.32(6) & 11.0 & 1.0(4)$\times$10$^\text{2}$ & 1.8(8) & -- & -- & -- & 26.6 & O01 & 30 & 4.3 \\
J1745$-$3040 & B1742$-$30 & 358.6 & $-$1.0 & 367.44 & 88.16(9) & 20.1 & 1.6(7)$\times$10$^\text{2}$ & 1.4(6) & -- & -- & -- & 37.9 & O04 & 80 & 7.3 \\
J1748$-$1300 & B1745$-$12 & 14.0 & 7.7 & 394.13 & 99.46(10) & 9.8 & 1.0(4)$\times$10$^\text{2}$ & 1.4(6) & -- & -- & -- & 50.8 & O01 & 30 & 5.3 \\
J1751$-$4657 & B1747$-$46 & 345.0 & $-$10.2 & 742.35 & 20.55(6) & 251.2 & 3.4(1.0)$\times$10$^\text{2}$ & 13(4) & 17.203(24) & 17.98(5) & 26.0 & 19.0 & O09 & 41 & 13.4 \\
J1752$-$2806 & B1749$-$28 & 1.5 & $-$1.0 & 562.57 & 50.40(5) & 970.7 & 3.6(1.4)$\times$10$^\text{3}$ & 1.7(7)e+02 & 94.83(9) & 96.05(10) & 19.5 & 10.2 & O04 & 59 & 3.7 \\
J1801$-$0357 & B1758$-$03 & 23.6 & 9.3 & 921.49 & 120.26(17) & 13.0 & 1.4(6)$\times$10$^\text{2}$ & 1.8(7) & -- & -- & -- & 30.2 & O05 & 30 & 7.6 \\
J1807$-$0847 & B1804$-$08 & 20.1 & 5.6 & 163.73 & 112.31(21) & 14.7 & 1.8(7)$\times$10$^\text{2}$ & 0.9(4) & -- & -- & -- & -- & O05 & 30 & 10.6 \\
J1819$-$1318 & -- & 17.5 & 0.8 & 1515.70 & 37.12(13) & 6.6 & 54(29) & 2.7(1.1) & -- & -- & -- & 31.6 & O06 & 30 & 6.8 \\
J1820$-$0427 & B1818$-$04 & 25.5 & 4.7 & 598.08 & 84.39(5) & 135.9 & 1.4(5)$\times$10$^\text{3}$ & 7.8(3.1) & 68.58(12) & 69.88(14) & 285.3 & 50.3 & O05 & 36 & 6.8 \\
J1822$-$4209 & -- & 351.9 & $-$12.8 & 456.51 & 72.03(7) & 13.3 & 50(16) & 0.63(22) & -- & -- & -- & 47.7 & O08 & 76 & 9.1 \\
J1823+0550 & B1821+05 & 35.0 & 8.9 & 752.91 & 66.83(6) & 40.9 & 1.6(7)$\times$10$^\text{2}$ & 10(4) & 140.99(13) & 142.62(14) & 17.6 & 8.9 & O07 & 39 & 14.7 \\
J1823$-$3106 & B1820$-$31 & 2.1 & $-$8.3 & 284.06 & 50.225(24) & 25.3 & 1.2(5)$\times$10$^\text{2}$ & 3.0(1.2) & 89.68(9) & 90.53(10) & 17.5 & 8.6 & O08 & 30 & 13.2 \\
J1825$-$0935 & B1822$-$09 & 21.4 & 1.3 & 769.02 & 19.34(7) & 24.0 & 2.1(8)$\times$10$^\text{2}$ & 9(4) & 67.61(21) & 68.82(22) & 25.3 & 12.8 & O05 & 14 & 15.9 \\
J1834$-$0010 & B1831$-$00 & 30.8 & 3.7 & 520.95 & 89.58(20) & 7.6 & 1.0(5)$\times$10$^\text{2}$ & 1.2(6) & -- & -- & -- & -- & O06 & 30 & 13.4 \\
J1834$-$0426 & B1831$-$04 & 27.0 & 1.7 & 290.11 & 79.448(25) & 82.6 & 9(4)$\times$10$^\text{2}$ & 7.7(3.1) & 92.35(13) & 93.65(14) & 85.3 & 20.7 & O05 & 36 & 6.7 \\
J1839$-$0627 & -- & 25.8 & $-$0.3 & 484.91 & 92.33(10) & 8.1 & 8(4)e+01 & 1.5(7) & -- & -- & -- & 30.5 & O06 & 30 & 7.2 \\
J1841+0912 & B1839+09 & 40.1 & 6.3 & 381.32 & 49.143(33) & 38.0 & 9(4)e+01 & 4.3(1.7) & 50.98(28) & 52.63(29) & 16.8 & 5.7 & O07 & 60 & 11.3 \\
J1847$-$0402 & B1844$-$04 & 28.9 & $-$0.9 & 597.81 & 141.7(7) & 22.9 & 4.6(1.8)$\times$10$^\text{2}$ & 1.5(7) & -- & -- & -- & 385.9 & O06 & 30 & 9.2 \\
J1848$-$1952 & B1845$-$19 & 14.8 & $-$8.3 & 4308.19 & 18.0(4) & 13.8 & 9(4)e+01 & 7.0(2.8) & 3.63(20) & 4.75(20)$^\dagger$ & 153.8 & 76.5 & O06 & 30 & 7.2 \\
J1849$-$0636 & B1846$-$06 & 26.8 & $-$2.5 & 1451.36 & 146.7(2.1) & 13.4 & 1.7(7)$\times$10$^\text{2}$ & 1.2(6) & -- & -- & -- & -- & O06 & 30 & 6.7 \\
J1851$-$0053 & -- & 32.1 & $-$0.3 & 1409.07 & 24.49(23) & 6.5 & 39(25) & 1.8(8) & -- & -- & -- & -- & O06 & 30 & 12.3 \\
J1900$-$2600 & B1857$-$26 & 10.3 & $-$13.5 & 612.21 & 37.89(5) & 203.6 & 5.7(1.7)$\times$10$^\text{2}$ & 19(6) & $-$8.51(9) & $-$7.62(11) & 48.9 & 11.4 & O08 & 30 & 14.7 \\
J1901+0156 & B1859+01 & 35.8 & $-$1.4 & 288.22 & 105.459(30) & 18.7 & 9(4)e+01 & 1.2(5) & $-$33.52(26) & $-$31.95(28)$^\dagger$ & -- & 17.8 & O07 & 65 & 17.3 \\
J1901$-$0906 & -- & 26.0 & $-$6.4 & 1781.93 & 72.47(15) & 6.6 & 27(14) & 1.1(5) & -- & -- & -- & 39.8 & O06 & 80 & 6.9 \\
J1909+0007 & B1907+00 & 35.1 & $-$4.0 & 1016.95 & 112.58(10) & 7.8 & 77(35) & 1.7(7) & -- & -- & -- & 47.8 & O05 & 36 & 10.3 \\
J1909+0254 & B1907+02 & 37.6 & $-$2.7 & 989.84 & 171.80(8) & 11.8 & 1.2(5)$\times$10$^\text{2}$ & 2.2(9) & -- & -- & -- & 34.3 & O05 & 36 & 10.3 \\
J1909+1102 & B1907+10 & 44.8 & 1.0 & 283.64 & 150.012(35) & 61.4 & 3.9(1.6)$\times$10$^\text{2}$ & 1.7(7) & 551.63(18) & 553.30(21) & 166.0 & 47.9 & O07 & 80 & 9.3 \\
J1910$-$0309 & B1907$-$03 & 32.3 & $-$5.7 & 504.61 & 206.09(23) & 16.1 & 1.6(7)$\times$10$^\text{2}$ & 1.2(6) & -- & -- & -- & 100.2 & O05 & 36 & 11.5 \\
J1913$-$0440 & B1911$-$04 & 31.3 & $-$7.1 & 825.94 & 89.34(7) & 160.8 & 5.8(2.3)$\times$10$^\text{2}$ & 24(10) & 3.47(7) & 4.68(12) & 36.7 & 13.8 & O06 & 30 & 8.6 \\
J1916+0951 & B1914+09 & 44.6 & $-$1.0 & 270.26 & 61.023(23) & 28.8 & 1.1(5)$\times$10$^\text{2}$ & 2.0(8) & 97.62(17) & 99.28(21) & 25.0 & 11.0 & O07 & 80 & 10.3 \\
J1917+1353 & B1915+13 & 48.3 & 0.6 & 194.64 & 94.718(17) & 60.5 & 1.8(7)$\times$10$^\text{2}$ & 1.8(7) & 227.54(18) & 229.26(21) & 45.5 & 15.3 & O07 & 80 & 7.1 \\
J1919+0021 & B1917+00 & 36.5 & $-$6.2 & 1272.27 & 90.12(13) & 11.3 & 47(23) & 2.2(9) & -- & -- & -- & 42.4 & O06 & 30 & 13.6 \\
J1921+1948 & B1918+19 & 53.9 & 2.7 & 821.04 & 153.81(9) & 20.5 & 83(34) & 0.9(4) & 161.7(4) & 163.5(5) & -- & 69.8 & O07 & 80 & 5.2 \\
J1921+2153 & B1919+21 & 55.8 & 3.5 & 1337.30 & 12.38(11) & 539.4 & 1.8(7)$\times$10$^\text{2}$ & 9(4) & $-$17.638(19) & $-$15.79(13) & 41.8 & 27.5 & O07 & 80 & 6.2 \\
J1922+2110 & B1920+21 & 55.3 & 2.9 & 1077.92 & 217.07(9) & 20.0 & 71(29) & 1.6(7) & -- & -- & 149.9 & 24.0 & O07 & 80 & 5.9 \\
J1932+1059 & B1929+10 & 47.4 & $-$3.9 & 226.52 & 3.173(19) & 304.3 & 4.8(1.9)$\times$10$^\text{2}$ & 14(6) & $-$7.573(8) & $-$5.90(12) & 17.1 & 4.0 & O07 & 75 & 9.9 \\
J1935+1616 & B1933+16 & 52.4 & $-$2.1 & 358.75 & 158.666(31) & 64.2 & 2.9(1.2)$\times$10$^\text{2}$ & 4.3(1.7) & $-$2.04(26) & $-$0.31(30) & 46.4 & 15.1 & O07 & 40 & 3.5 \\
J1941+0121 & -- & 39.9 & $-$10.4 & 217.32 & 52.21(15) & 9.4 & 71(23) & 0.64(25) & -- & -- & -- & 24.8 & O11 & 30 & 3.1 \\
J1943$-$1237 & B1940$-$12 & 27.3 & $-$17.2 & 972.43 & 28.97(8) & 38.6 & 1.5(5)$\times$10$^\text{2}$ & 9.6(2.9) & -- & -- & 24.0 & 12.1 & O06 & 30 & 2.2 \\
J1944$-$1750 & B1941$-$17 & 22.3 & $-$19.4 & 841.16 & 56.27(7) & 11.4 & 69(23) & 2.6(8) & -- & -- & 47.1 & 20.3 & O06 & 30 & 5.4 \\
J1946+1805 & B1944+17 & 55.3 & $-$3.5 & 440.62 & 16.10(7) & 19.1 & 68(28) & 0.9(4) & $-$45.91(30) & $-$44.14(33) & -- & 46.1 & O07 & 60 & 5.9 \\
J1946$-$2913 & B1943$-$29 & 11.1 & $-$24.1 & 959.45 & 44.26(8) & 7.8 & 34(14) & 3.0(9) & 3.82(15) & 4.69(16) & 30.0 & 9.7 & O08 & 30 & 13.7 \\
J1949$-$2524 & B1946$-$25 & 15.3 & $-$23.4 & 957.62 & 23.02(8) & 5.3 & 12(11) & 0.83(30) & -- & -- & -- & 15.0 & O06 & 30 & 12.8 \\
J2043+2740 & -- & 70.6 & $-$9.2 & 96.13 & 21.041(8) & 11.9 & 47(20) & 1.0(4) & $-$96.88(15) & $-$95.07(18) & -- & 4.5 & O14 & 30 & 10.0 \\
J2046$-$0421 & B2043$-$04 & 42.7 & $-$27.4 & 1546.94 & 35.89(13) & 130.9 & 78(24) & 4.2(1.3) & $-$2.75(6) & $-$1.62(16)$^\dagger$ & 48.5 & 24.0 & O13 & 59 & 11.5 \\
J2046+1540 & B2044+15 & 61.1 & $-$16.8 & 1138.29 & 39.76(10) & 15.2 & 34(12) & 1.8(6) & $-$90.56(24) & $-$89.01(26) & 68.6 & 14.2 & O14 & 30 & 4.4 \\
J2048$-$1616 & B2045$-$16 & 30.5 & $-$33.1 & 1961.58 & 11.55(17) & 430.5 & 2.6(8)$\times$10$^\text{2}$ & 9.1(2.7) & $-$10.450(11) & $-$9.42(14) & 112.0 & 95.7 & O13 & 71 & 8.4 \\
J2053$-$7200 & B2048$-$72 & 321.9 & $-$35.0 & 341.34 & 17.610(29) & 70.7 & 2.1(6)$\times$10$^\text{2}$ & 5.2(1.6) & 18.74(5) & 19.37(6)$^\dagger$ & 42.8 & 5.8 & O12 & 30 & 3.1 \\
J2055+2209 & B2053+21 & 67.8 & $-$14.7 & 815.18 & 36.40(7) & 13.3 & 28(10) & 1.7(5) & $-$90.08(19) & $-$88.39(20) & 34.8 & 21.6 & O14 & 30 & 4.2 \\
J2116+1414 & B2113+14 & 64.5 & $-$23.4 & 440.15 & 56.16(4) & 9.4 & 32(11) & 0.60(20) & -- & -- & -- & 17.0 & O14 & 30 & 6.2 \\
J2139+2242 & -- & 75.3 & $-$21.9 & 1083.51 & 43.87(27) & 8.0 & 28(10) & 0.51(18) & -- & -- & -- & 90.0 & O14 & 26 & 9.9 \\
J2155$-$3118 & B2152$-$31 & 15.8 & $-$51.6 & 1030.00 & 14.90(9) & 83.6 & 68(21) & 3.2(1.0) & 13.44(4) & 14.56(10)$^\dagger$ & 37.4 & 24.6 & O15 & 54 & 12.8 \\
J2253+1516 & -- & 85.3 & $-$38.8 & 792.24 & 29.28(7) & 8.4 & 8(5) & 0.54(17) & -- & -- & -- & 15.4 & B05 & 42 & 6.6 \\
J2324$-$6054 & B2321$-$61 & 320.4 & $-$53.2 & 2347.49 & 14.95(21) & 10.9 & 10(4) & 0.43(13) & -- & -- & -- & 99.6 & B09 & 74 & 11.4 \\
J2325$-$0530 & -- & 75.6 & $-$60.2 & 868.74 & 14.93(7) & 12.3 & 15(5) & 1.17(35) & -- & -- & 17.9 & 10.2 & B04 & 80 & 9.6 \\
J2330$-$2005 & B2327$-$20 & 49.4 & $-$70.2 & 1643.62 & 8.38(14) & 192.0 & 1.3(4)$\times$10$^\text{2}$ & 8.8(2.6) & 9.220(27) & 9.97(11) & 42.4 & 25.4 & B04 & 30 & 8.8 \\
J2336$-$01 & -- & 84.4 & $-$59.0 & 1029.80 & 19.55(9) & 23.6 & 40(12) & 1.05(32) & -- & -- & 64.8 & 36.8 & B04 & 80 & 12.6 \\
J2346$-$0609 & -- & 83.8 & $-$64.0 & 1181.46 & 22.60(10) & 6.5 & 13(6) & 0.78(24) & -- & -- & -- & 14.7 & B04 & 30 & 11.6 \\
J2354$-$2250 & -- & 48.1 & $-$76.4 & 558.00 & 9.97(5) & 20.2 & 12(5) & 0.78(24) & 9.92(20) & 10.72(23)$^\dagger$ & 27.2 & 10.6 & B07 & 63 & 9.7 \\

\end{longtable}

\end{landscape}

\twocolumn

\end{document}